\documentclass[twocolumn,aps,pra,amsmath,amssymb]{revtex4-1}
\usepackage[latin9]{inputenc}
\usepackage{amsmath}
\usepackage{amssymb}
\usepackage{graphicx}
\usepackage{esint}

\makeatletter

\newcommand*\LyXThinSpace{\,\hspace{0pt}}

\usepackage{epsfig}
\usepackage{bm}
\usepackage{dcolumn}
\usepackage{color}

\makeatother

\begin{document}
\title{Fermi spin polaron and dissipative Fermi-polaron Rabi dynamics}
\author{Hui Hu$^{1}$ and Xia-Ji Liu$^{1,2}$}
\affiliation{$^{1}$Centre for Quantum Technology Theory, Swinburne University
of Technology, Melbourne, Victoria 3122, Australia}
\affiliation{$^{2}$Kavli Institute for Theoretical Physics, UC Santa Barbara,
USA}
\date{\today}
\begin{abstract}
We consider a spin impurity with multiple energy levels moving in
a non-interacting Fermi sea, and theoretically solve this Fermi spin
polaron problem at nonzero temperature by using a non-self-consistent
many-body $T$-matrix theory. We focus on the simplest case with spin
half, where the two energy states of the impurity are coupled by a
Rabi flip term. At small Rabi coupling, the impurity exhibits damped
Rabi oscillations, where the decoherence is caused by the interaction
with the Fermi sea, as recently reported in Fermi polaron experiments
with ultracold atoms. We investigate the dependence of Rabi oscillations
on the Rabi coupling strength and examine the additional nonlinear
damping due to large Rabi coupling. At finite temperature and at nonzero
impurity concentration, the impurity can acquire a pronounced momentum
distribution. We show that the momentum/thermal average can sizably
reduce the visibility of Rabi oscillations. We compare our theoretical
predictions to the recent experimental data and find a good agreement
without any adjustable parameter.
\end{abstract}
\maketitle

\section{Introduction}

Quantum impurity interacting with a many-body environment is a long-lasting
research topic in the modern physics \cite{Alexandrov2010}. The earliest
study can be traced back to the seminal work by Lev Landau in 1933
\cite{Landau1933}, which led to the fundamental concept of quasiparticles.
Over the last fifteen years, this research topic has received renewed
interest, due to rapid advances in ultracold atomic physics \cite{Bloch2008,Chin2010,Hu2022AB}.
In particular, the dynamics of a quantum impurity immersed in a non-interacting
Fermi sea, namely, Fermi polaron, has been systematically explored
both experimentally and theoretically \cite{Massignan2014,Lan2014,Schmidt2018,Mistakidis2019,Seetharam2021}.
A convenient experimental setup is the use of a highly imbalanced
two-component Fermi-Fermi mixture, where minority atoms in a hyperfine
state can be well treated as isolated, uncorrelated impurities. For
such a system, quasiparticle properties of Fermi polarons, including
the ground-state attractive polaron and the excited branch of repulsive
polaron, have been well characterized experimentally by the radio-frequency
(rf) spectroscopy \cite{Schirotzek2009,Zhang2012,Kohstall2012,Koschorreck2012,Scazza2017,Zan2019},
Ramsey interferometry \cite{Cetina2016}, Rabi oscillation \cite{Kohstall2012,Scazza2017,DarkwahOppong2019},
and most recently by Raman spectroscopy \cite{Ness2020}. Theoretically,
an exactly solvable polaron model with a heavy impurity and a Bardeen-Cooper-Schrieffer
(BCS) superfluid environment has also been constructed \cite{Wang2022PRL,Wang2022PRA,Wang2023AB},
clarifying several salient features of Fermi polarons in a rigorous
way.

In principle, quantum impurity can have internal degrees of freedom
and can occupy multiple energy levels. For example, molecule impurity
can be trapped inside a nanodroplet of superfluid helium, forming
the so-called angulon quasiparticle \cite{Schmidt2015}. The rotational
degree of freedom of the molecule can be affected by the many-body
environment of helium droplet, as evidenced by a larger effective
moment of inertia. This effect is similar to the renormalization of
the effective mass for impurity observed in Fermi polarons \cite{Massignan2014}.
In highly imbalanced Fermi-Fermi mixtures, it is also feasible to
coherently transfer minority atoms to another hyperfine state, by
using an always-on rf field \cite{Knap2013}. Thus, impurity atoms
can occupy two different hyperfine states and acquire a pseduo-spin
degree of freedom. Indeed, in recent Rabi dynamics experiments for
Fermi polarons \cite{Kohstall2012,Scazza2017,DarkwahOppong2019},
Rabi oscillation between the two hyperfine states is driven by the
rf field with reasonably small coupling strength in the linear response
regime, where polaron properties are assumed to be unchanged by Rabi
coupling.

In this work, we investigate in detail the Fermi spin polaron with
a mobile \emph{spinor} impurity, with the purpose of better understanding
the Rabi dynamics of Fermi polarons. We are specifically interested
in the dependence of quasiparticle properties of Fermi spin polarons
on the Rabi coupling strength, which is less considered in earlier
theoretical analyses on dissipative Rabi dynamics \cite{Knap2013,Parish2016,Adlong2021}
(For an exception, see Ref. \cite{Adlong2020}, where a state-of-the-art
simulation of Rabi oscillations is presented). This dependence is
crucial to examine the small Rabi coupling assumption adopted in recent
experimental measurements \cite{Kohstall2012,Scazza2017,DarkwahOppong2019}.

Our theoretical investigation is based on a non-self-consistent many-body
$T$-matrix theory of Fermi polarons \cite{Combescot2007,Hu2018,Wang2019,Mulkerin2019,Tajima2019,Tajima2021,Hu2022CrossoverPolaron,Hu2022,Hu2023ABpwave},
extended to the case of a spinor impurity. In the spinless case of
a structureless impurity, such a many-body $T$-matrix approach is
fully equivalent to Chevy's variational ansatz \cite{Chevy2006,Cui2010,Parish2013,Hu2023AB},
including its finite temperature extension \cite{Liu2019}. This approach
is particularly useful for a mobile impurity, whose recoil energy
suppresses multiple particle-hole excitations near the Fermi surface
of the many-body environment. Our results are therefore complement
to the two earlier studies \cite{Knap2013,Adlong2021}, which considered
the heavy impurity limit using either a spin-model with an Ohmic bath
\cite{Knap2013} or the functional determinant approach \cite{Adlong2021}. 

Our many-body $T$-matirx theory is also convenient to investigate
the finite-momentum effect of polarons, which arises due to the nonzero
temperature and the finite impurity concentration. This effect is
not emphasized in a recent Rabi dynamics study based on the finite-temperature
variation approach \cite{Adlong2020}, but is found to be important
for understanding the measured rf spectroscopy \cite{Hu2022}. We
find that the visibility of Rabi oscillations can be sizably reduced
by the momentum average due to the thermal momentum distribution of
polarons.

The rest of the paper is organized as follows. In the next section
(Sec. II), we present the non-self-consistent many-body $T$-matrix
theory for Fermi spin polarons at finite temperature. In Sec. III,
we discuss in detail the quasiparticle properties of spin polarons,
such as self-energy, spectral function and polaron energies, as a
function of the Rabi coupling strength. We emphasize the nonlinear
effect arising from large Rabi coupling. In Sec. VI, we first compare
our theoretical predictions with the experimental data on Rabi oscillation
and show that there is a good agreement, without any free fitting
parameters. We then examine the effect of the momentum average and
the nonlinear dependence of Rabi oscillations on large Rabi coupling
strength. Finally, we give a brief summary and outlook in Sec. V.

\section{The non-self-consistent many-body $T$-matrix theory}

\subsection{The model Hamiltonian}

According to the recent experiments on dissipative Rabi dynamics \cite{Kohstall2012,Scazza2017,DarkwahOppong2019},
we consider a spin-1/2 impurity of mass $m_{I}$ that has two hyperfine
energy levels (i.e., $\sigma=\uparrow,\downarrow$), described by
the single-particle model Hamiltonian \cite{Knap2013,Adlong2021},
\begin{equation}
\mathcal{H}_{I}=\sum_{\mathbf{p}\sigma}\epsilon_{\mathbf{p\sigma}}^{(I)}d_{\mathbf{p}\sigma}^{\dagger}d_{\mathbf{p}\sigma}+\frac{\Omega}{2}\sum_{\mathbf{p}}\left(d_{\mathbf{p}\uparrow}^{\dagger}d_{\mathbf{p}\downarrow}+d_{\mathbf{p}\downarrow}^{\dagger}d_{\mathbf{p}\uparrow}\right),
\end{equation}
where $d_{\mathbf{p\sigma}}^{\dagger}$ and $d_{\mathbf{p}\sigma}$
are the creation and annihilation field operators for the impurity
with momentum $\mathbf{p}$ in the spin-up ($\sigma=\uparrow$) and
spin-down ($\sigma=\downarrow$) states that have the dispersion relations
$\epsilon_{\mathbf{p\uparrow}}^{(I)}=\epsilon_{\mathbf{p}}^{(I)}\equiv\hbar^{2}\mathbf{p}^{2}/(2m_{I})$
and $\epsilon_{\mathbf{p\downarrow}}^{(I)}\equiv\epsilon_{\mathbf{p}}^{(I)}+\Delta$
respectively, $\Delta$ is the detuning, and $\Omega$ is the Rabi
coupling strength.

For a spin-1/2 impurity, its non-interacting thermal Green function
is a 2 by 2 matrix $\mathbf{G}_{0}(\mathcal{P})$,
\begin{equation}
\left[\begin{array}{cc}
G_{11}^{(0)} & G_{12}^{(0)}\\
G_{21}^{(0)} & G_{22}^{(0)}
\end{array}\right]=\left[\begin{array}{cc}
i\omega_{p}-\epsilon_{\mathbf{p}}^{(I)} & -\Omega/2\\
-\Omega/2 & i\omega_{p}-\epsilon_{\mathbf{p}}^{(I)}-\Delta
\end{array}\right]^{-1},
\end{equation}
where we have used the short-hand notation $\mathcal{P}\equiv(\mathbf{p},i\omega_{p})$
with fermionic Matsubara frequency $\omega_{p}=(2p+1)\pi k_{B}T$
at temperature $T$ and integer $p=0,\pm1,\pm2,\cdots$. By diagonalizing
the matrix, we find two energy levels, $E_{\mathbf{p}}^{(\pm)}=(\epsilon_{\mathbf{p}}^{(I)}+\Delta/2)\pm\sqrt{\Delta^{2}+\Omega^{2}}/2$.
The associated amplitudes (i.e., wavefunctions) are given by,
\begin{eqnarray}
u^{2} & = & \frac{1}{2}\left[1+\frac{\Delta}{\sqrt{\Delta^{2}+\Omega^{2}}}\right],\\
v^{2} & = & \frac{1}{2}\left[1-\frac{\Delta}{\sqrt{\Delta^{2}+\Omega^{2}}}\right],\\
uv & = & \frac{1}{2}\frac{\Omega}{\sqrt{\Delta^{2}+\Omega^{2}}}.
\end{eqnarray}
The non-interacting impurity Green function can then be conveniently
written as,
\begin{equation}
\left[\begin{array}{cc}
G_{11}^{(0)} & G_{12}^{(0)}\\
G_{21}^{(0)} & G_{22}^{(0)}
\end{array}\right]=\frac{\left[\begin{array}{cc}
v^{2} & uv\\
uv & u^{2}
\end{array}\right]}{i\omega_{p}-E_{\mathbf{p}}^{(+)}}+\frac{\left[\begin{array}{cc}
u^{2} & -uv\\
-uv & v^{2}
\end{array}\right]}{i\omega_{p}-E_{\mathbf{p}}^{(-)}}.\label{eq:impurityGF0}
\end{equation}

The impurity is moving in and interacting with an ideal Fermi sea
of fermionic atoms of mass $m$ described by $\sum_{\mathbf{k}}\epsilon_{\mathbf{k}}c_{\mathbf{k}}^{\dagger}c_{\mathbf{k}}$,
where $c_{\mathbf{k}}^{\dagger}$ and $c_{\mathbf{k}}$ are the creation
and annihilation field operators for fermionic atoms with momentum
$\mathbf{k}$ and single-particle dispersion relation $\epsilon_{\mathbf{k}}=\hbar^{2}\mathbf{k}^{2}/(2m)$.
The total model Hamiltonian then takes the form,

\begin{equation}
\mathcal{H}=\mathcal{H}_{I}+\sum_{\sigma}\frac{g_{\sigma}}{V}\sum_{\mathbf{kpq}}c_{\mathbf{k}}^{\dagger}d_{\mathbf{q}-\mathbf{k}\sigma}^{\dagger}d_{\mathbf{\mathbf{q}-p}\sigma}c_{\mathbf{p}}+\sum_{\mathbf{k}}\epsilon_{\mathbf{k}}c_{\mathbf{k}}^{\dagger}c_{\mathbf{k}},
\end{equation}
where $V$ is the system volume and the middle term describes the
$s$-wave contact interactions between the impurity and the Fermi
bath with bare interaction strengths $g_{\sigma}$, which are to be
regularized via the relation, 
\begin{equation}
\frac{1}{g_{\sigma}}=\frac{m_{r}}{2\pi\hbar^{2}a_{\sigma}}-\frac{1}{V}\sum_{\mathbf{k}}\frac{2m_{r}}{\hbar^{2}\mathbf{k}^{2}}.\label{eq:gs}
\end{equation}
Here $a_{\sigma}$ ($\sigma=\uparrow,\downarrow$ or interchangeably
$\sigma=1,2$) is the $s$-wave impurity-bath scattering length, $m_{r}\equiv mm_{I}/(m+m_{I})$
is the reduced mass. Throughout the work, we always take $m_{I}=m$,
so $m_{r}=m/2$. The density ($n$) or the total number ($N=nV$)
of fermionic atoms in the Fermi sea can be tuned by adjusting the
temperature-dependent chemical potential $\mu(T)$. We often measure
the single-particle energy of atoms from the chemical potential and
therefore define $\xi_{\mathbf{k}}\equiv\epsilon_{\mathbf{k}}-\mu$.
Hereafter, for clarity we will suppress the volume $V$ in expressions,
so the summation over the momentum $\sum_{\mathbf{k}}$ in the later
equations should be understood as $\sum_{\mathbf{k}}=(1/V)\sum_{\mathbf{k}}=\int d\mathbf{k}/(2\pi)^{3}$.

\subsection{The diagrammatic theory}

We use the non-self-consistent many-body $T$-matrix theory \cite{Combescot2007,Hu2022}
to solve the Fermi spin polaron problem, within which the motion of
the impurity can be described by a series of \emph{ladder} diagrams
that take into account the successive forward scatterings between
the impurity and the atoms in the Fermi bath. By summing up the infinitely
many ladder diagrams, as detailed in Appendix A, we find the two-particle
vertex function, which takes the following 2 by 2 matrix form,
\begin{equation}
\left[\begin{array}{cc}
\Gamma_{11} & \Gamma_{12}\\
\Gamma_{21} & \Gamma_{22}
\end{array}\right]=\left[\begin{array}{cc}
1/g_{1}+\tilde{\chi}_{11}\left(\mathcal{Q}\right) & \tilde{\chi}_{12}\left(\mathcal{Q}\right)\\
\tilde{\chi}_{21}\left(\mathcal{Q}\right) & 1/g_{2}+\tilde{\chi}_{22}\left(\mathcal{Q}\right)
\end{array}\right]^{-1},\label{eq:vertexfunction}
\end{equation}
where $\mathcal{Q}\equiv(\mathbf{q},iv_{q})$ is the short-hand notation
of the four-dimensional momentum with bosonic Matsubara frequency
$\nu_{q}=2q\pi k_{B}T$ and integer $q=0,\pm1,\pm2,\cdots$, and the
various pair propagators $\tilde{\chi}_{ij}$ ($i,j=1,2$) are given
by,
\begin{equation}
\tilde{\chi}_{ij}\left(\mathbf{\mathcal{Q}}\right)=\sum_{\mathbf{k}}k_{B}T\sum_{i\omega_{k}}\mathcal{G}\left(\mathcal{K}\right)G_{ij}^{(0)}\left(\mathcal{Q}-\mathcal{K}\right).\label{eq:kappa2p}
\end{equation}
Here we have introduced $\mathcal{K}\equiv(\mathbf{k},i\omega_{k})$
with fermionic Matsubara frequency $\omega_{k}=(2k+1)\pi k_{B}T$
and integer $k=0,\pm1,\pm2,\cdots$, and 
\begin{equation}
\mathcal{G}(\mathcal{K})=\frac{1}{i\omega_{k}-\xi_{\mathbf{k}}}=\frac{1}{i\omega_{k}-\epsilon_{\mathbf{k}}+\mu}
\end{equation}
is the thermal Green function for non-interacting fermionic atoms
in the Fermi bath. The summation over the fermionic Matsubara frequency
in Eq. (\ref{eq:kappa2p}) is easy to carry out. In the single impurity
limit, we find that \cite{Hu2022}, 
\begin{eqnarray}
\tilde{\chi}_{11} & = & \sum_{\mathbf{k}}\left[\frac{v^{2}\left(f\left(\xi_{\mathbf{k}}\right)-1\right)}{i\nu_{q}-E_{\mathbf{q}-\mathbf{k}}^{(+)}-\xi_{\mathbf{k}}}+\frac{u^{2}\left(f\left(\xi_{\mathbf{k}}\right)-1\right)}{i\nu_{q}-E_{\mathbf{q}-\mathbf{k}}^{(-)}-\xi_{\mathbf{k}}}\right],\\
\tilde{\chi}_{12} & = & \sum_{\mathbf{k}}\left[\frac{uv\left(f\left(\xi_{\mathbf{k}}\right)-1\right)}{i\nu_{q}-E_{\mathbf{q}-\mathbf{k}}^{(+)}-\xi_{\mathbf{k}}}-\frac{uv\left(f\left(\xi_{\mathbf{k}}\right)-1\right)}{i\nu_{q}-E_{\mathbf{q}-\mathbf{k}}^{(-)}-\xi_{\mathbf{k}}}\right],
\end{eqnarray}
$\tilde{\chi}_{21}=\tilde{\chi}_{12}$, and $\tilde{\chi}_{22}$ can
be obtained from $\tilde{\chi}_{11}$ by exchanging the factor $u^{2}$
with $v^{2}$ in the square bracket. The function $f(x)=1/[e^{x/(k_{B}T)}+1]$
is the Fermi-Dirac distribution at temperature $T$. It is readily
seen that the integral in both $\tilde{\chi}_{11}$ and $\tilde{\chi}_{22}$
has an ultraviolet divergence at large momentum. This divergence is
due to the use of the $s$-wave contact interactions and can be exactly
compensated by the counter term in the regularization relation Eq.
(\ref{eq:gs}), i.e., $\sum_{\mathbf{k}}2m_{r}/(\hbar^{2}\mathbf{k}^{2})$.
Therefore, it is convenient to introduce $\chi_{11}\equiv1/g_{1}+\tilde{\chi}_{11}$
and $\chi_{22}\equiv1/g_{2}+\tilde{\chi}_{22}$, and rewrite $\chi_{12}\equiv\tilde{\chi}_{12}$
and $\chi_{21}\equiv\tilde{\chi}_{21}$. We will still refer to $\chi_{ij}(\mathcal{Q})$
as the pair propagators, without any confusion.

The integrals in $\chi_{ij}(\mathcal{Q})$ can be categorized into
two types \cite{Hu2022}. The first is the two-body part, which can
be analytically evaluated by using,
\begin{equation}
\sum_{\mathbf{k}}\left[\frac{1}{\Omega-\epsilon_{\mathbf{q}-\mathbf{k}}^{(I)}-\xi_{\mathbf{k}}}+\frac{2m_{r}}{\hbar^{2}\mathbf{k}^{2}}\right]=-\frac{i\left(2m_{r}\right)^{\frac{3}{2}}\sqrt{\Omega-\zeta_{\mathbf{q}}}}{4\pi\hbar^{3}}
\end{equation}
for any complex frequency $\Omega$. Here $\zeta_{\mathbf{q}}\equiv\hbar^{2}\mathbf{q}^{2}/[2(m+m_{I})]-\mu$
is the center-of-mass kinetic energy measured from the chemical potential.
Another is the many-body part, which takes the form,
\begin{equation}
\chi_{\textrm{eff}}\left(\mathbf{q},\Omega\right)=\sum_{\mathbf{k}}\frac{f\left(\xi_{\mathbf{k}}\right)}{\Omega-\epsilon_{\mathbf{q}-\mathbf{k}}^{(I)}-\xi_{\mathbf{k}}},
\end{equation}
and can be numerically calculated in a very efficient way, as discussed
in detail in our recent work (see, i.e., Appendix A of Ref. \cite{Hu2022}).
By defining two constants $\gamma_{\pm}=[\Delta\pm\sqrt{\Delta^{2}+\Omega^{2}}]/2$
and rewriting $E_{\mathbf{p}}^{(\pm)}=\epsilon_{\mathbf{p}}^{(I)}+\gamma_{\pm}$,
it is then easy to check that ($\mathcal{Q}=(\mathbf{q},i\nu_{q})\equiv(\mathbf{q},\Omega)$),\begin{widetext}
\begin{eqnarray}
\chi_{11}\left(\mathcal{Q}\right) & = & \frac{m_{r}}{2\pi\hbar^{2}a_{1}}+\frac{im_{r}^{3/2}}{\sqrt{2}\pi\hbar^{3}}\left[v^{2}\sqrt{\Omega-\gamma_{+}-\zeta_{\mathbf{q}}}+u^{2}\sqrt{\Omega-\gamma_{-}-\zeta_{\mathbf{q}}}\right]+v^{2}\chi_{\textrm{eff}}\left(\mathbf{q},\Omega-\gamma_{+}\right)+u^{2}\chi_{\textrm{eff}}\left(\mathbf{q},\Omega-\gamma_{-}\right),\label{eq:kappa11}\\
\chi_{12}\left(\mathcal{Q}\right) & = & \frac{im_{r}^{3/2}}{\sqrt{2}\pi\hbar^{3}}uv\left[\sqrt{\Omega-\gamma_{+}-\zeta_{\mathbf{q}}}-\sqrt{\Omega-\gamma_{-}-\zeta_{\mathbf{q}}}\right]+uv\left[\chi_{\textrm{eff}}\left(\mathbf{q},\Omega-\gamma_{+}\right)-\chi_{\textrm{eff}}\left(\mathbf{q},\Omega-\gamma_{-}\right)\right],\label{eq:kappa12}\\
\chi_{22}\left(\mathcal{Q}\right) & = & \frac{m_{r}}{2\pi\hbar^{2}a_{2}}+\frac{im_{r}^{3/2}}{\sqrt{2}\pi\hbar^{3}}\left[u^{2}\sqrt{\Omega-\gamma_{+}-\zeta_{\mathbf{q}}}+v^{2}\sqrt{\Omega-\gamma_{-}-\zeta_{\mathbf{q}}}\right]+u^{2}\chi_{\textrm{eff}}\left(\mathbf{q},\Omega-\gamma_{+}\right)+v^{2}\chi_{\textrm{eff}}\left(\mathbf{q},\Omega-\gamma_{-}\right).\label{eq:kappa22}
\end{eqnarray}
\end{widetext}Here, for late convenience of taking analytic continuation
(i.e., $i\nu_{q}\rightarrow\Omega+i0^{+}$), we have explicitly set
$i\nu_{q}=\Omega$. The analytic continuation can then be performed
by simply adding $i0^{+}$ to $\Omega$. 

Once the pair propagators $\chi_{ij}(\mathcal{Q})$ are obtained,
we take the matrix inverse to find the 2 by 2 vertex function $\Gamma(\mathcal{Q})=\left[\chi(\mathcal{Q})\right]{}^{-1}$.
The 2 by 2 self-energy of the impurity $\Sigma_{ij}(\mathcal{P})$
can be obtained by winding back the out-going leg of the fermionic
field operator in the vertex function $\Gamma(\mathcal{Q})_{ij}$
and by connecting it with the in-coming leg of the fermionic field
operator \cite{Combescot2007,Hu2022}. Physically, this describes
the \emph{single} particle-hole excitation across the Fermi surface
of the bath \cite{Combescot2007,Hu2022CrossoverPolaron,Chevy2006}.
Thus, we have
\begin{equation}
\Sigma_{ij}\left(\mathcal{P}\right)=\sum_{\mathbf{q}}k_{B}T\sum_{i\nu_{q}}\Gamma_{ij}\left(\mathcal{Q}\right)\frac{1}{i\nu_{q}-i\omega_{p}-\xi_{\mathbf{q}-\mathbf{p}}}.
\end{equation}
The summation over the bosonic Matsubara frequency $\nu_{q}=2q\pi k_{B}T$
(with integer $q=0,\pm1,\pm2,\cdots$) can be easily carried out,
leading to \cite{Combescot2007,Hu2022},
\begin{equation}
\Sigma_{ij}\left(\mathbf{p},i\omega_{p}\right)=\sum_{\mathbf{q}}f\left(\xi_{\mathbf{q}-\mathbf{p}}\right)\Gamma_{ij}\left(\mathbf{q},i\omega_{p}+\xi_{\mathbf{q}-\mathbf{p}}\right).\label{eq:selfenergy}
\end{equation}

\subsection{Analytic continuation and numerical calculations}

We are interested in the \emph{retarded} impurity Green functions
given by the Dyson equation,

\begin{equation}
\mathbf{G}\left(\mathbf{p},\omega\right)=\left[\mathbf{G}_{0}^{-1}\left(\mathbf{p},\omega\right)-\mathbf{\Sigma}\left(\mathbf{p},\omega\right)\right]^{-1},\label{eq:impurityGF}
\end{equation}
and the related impurity spectral functions $\mathbf{A}(\mathbf{p},\omega)$,
\begin{equation}
\left[\begin{array}{cc}
A_{11} & A_{12}\\
A_{21} & A_{22}
\end{array}\right]=-\frac{1}{\pi}\textrm{Im}\left[\begin{array}{cc}
G_{11}\left(\mathbf{p},\omega\right) & G_{12}\left(\mathbf{p},\omega\right)\\
G_{21}\left(\mathbf{p},\omega\right) & G_{22}\left(\mathbf{p},\omega\right)
\end{array}\right].
\end{equation}
Here, $\mathbf{G}_{0}(\mathbf{p},\omega)$ and $\mathbf{\Sigma}(\mathbf{p},\omega)$
are given by Eq. (\ref{eq:impurityGF0}) and Eq. (\ref{eq:selfenergy}),
respectively, with the analytic continuation (i.e., $i\omega_{p}\rightarrow\omega+i0^{+}$)
explicitly performed. As mentioned earlier, this analytic continuation
can be trivially done by replacing $\Omega$ in Eqs. (\ref{eq:kappa11})-(\ref{eq:kappa22})
with $\Omega+i0^{+}$ for the calculations of the retarded pair propagators
$\chi_{ij}(\mathbf{q},\Omega)$. The consequent matrix inverse leads
to the retarded two-particle vertex functions $\Gamma_{ij}\left(\mathbf{q},\Omega\right)$,
which is to be used in Eq. (\ref{eq:selfenergy}). 

In our numerical calculations, we take the Fermi wave-vector $k_{F}\equiv(6\pi^{2}n)^{1/3}$
and Fermi energy $\varepsilon_{F}\equiv\hbar^{2}k_{F}^{2}/(2m)$ as
the units of the momentum (or wave-vector) and energy, respectively.
The temperature $T$ is accordingly measured in units of the Fermi
temperature $T_{F}=\varepsilon_{F}/k_{B}$. The choice of this natural
units amounts to setting $2m=\hbar=k_{B}=1$. As we take the equal
mass for the impurity and background fermionic atoms ($m=m_{I}$),
the reduced mass $m_{r}=1/4$. In Eqs. (\ref{eq:kappa11})-(\ref{eq:kappa22}),
we find that $m_{r}/(2\pi\hbar^{2}a_{i})=1/(8\pi a_{i})$ ($i=1,2$),
$m_{r}^{3/2}/(\sqrt{2}\pi\hbar^{3})=1/(8\sqrt{2}\pi)$ and $\zeta_{\mathbf{q}}=\mathbf{q}^{2}/2-\mu$.
The dimensionless expression of $\chi_{\textrm{eff}}(\mathbf{q},\Omega+i0^{+})$
can be found in Appendix A of Ref. \cite{Hu2022}.

\begin{figure}
\begin{centering}
\includegraphics[clip,width=0.48\textwidth]{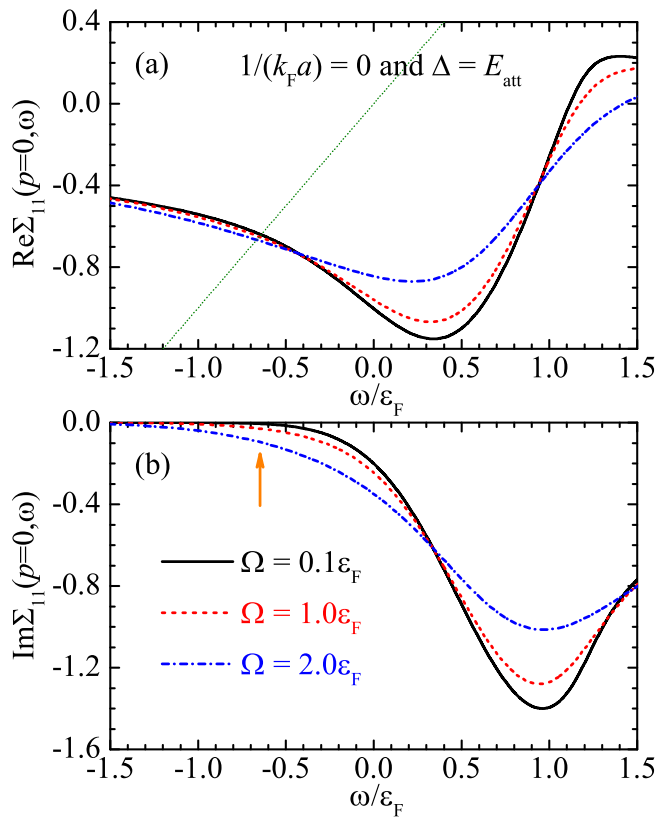}
\par\end{centering}
\caption{\label{fig1_seRabiUnitary} The real part (a) and imaginary part (b)
of the impurity self-energy $\Sigma_{11}(\mathbf{p},\omega)$ at zero
momentum ($\mathbf{p}=\mathbf{0}$) in the unitary limit $1/a=0$.
The self-energy is in units of of $\varepsilon_{F}$, where $\varepsilon_{F}\equiv\hbar^{2}k_{F}^{2}/(2m)$
and $k_{F}=(6\pi^{2}n)^{1/3}$ are the Fermi energy and Fermi wavevector,
respectively. The temperature is $T=0.2T_{F}=0.2\varepsilon_{F}/k_{B}$
and the detuning is equal to the attractive polaron energy (at zero
Rabi coupling), $\Delta=E_{\textrm{att}}$. We have consider three
characteristic Rabi couplings $\Omega=0.1\varepsilon_{F}$ (black
solid line), $1.0\varepsilon_{F}$ (red dashed line), and $2.0\varepsilon_{F}$
(blue dot-dashed line). The green dotted line in (a) shows the curve
$y=\omega$. The arrow in (b) points to the attractive polaron energy
$E_{\textrm{att}}\simeq-0.64\varepsilon_{F}$.}
\end{figure}

\section{Quasiparticle properties of Fermi spin polarons}

In the experiments on dissipative Rabi dynamics \cite{Kohstall2012,Scazza2017,DarkwahOppong2019},
the interaction between the spin-down impurity and the Fermi bath
($a_{\downarrow}$ or $a_{2}$) is typically small. For convenience,
we simply set $a_{2}=0^{-}$ and denote $a_{\uparrow}=a_{1}=a$. Therefore,
in Eq. (\ref{eq:kappa22}) $\chi_{22}(\mathcal{Q})\rightarrow\infty$.
By taking the matrix inverse, only the 11-component of the matrices,
such as the two-particle vertex function $\Gamma_{11}(\mathcal{Q})=\chi_{11}^{-1}(\mathcal{Q})$
and the self-energy $\Sigma_{11}(\mathbf{p},\omega)$, is nonzero.
We note that, the case with a small but positive scattering length
$a_{2}>0$ might be useful to understand the \emph{residual} final-state
effect in the rf-spectroscopy \cite{Schirotzek2009,Zan2019} or Raman
spectroscopy \cite{Ness2020}, and could be addressed in future studies.

\begin{figure}
\begin{centering}
\includegraphics[width=0.48\textwidth]{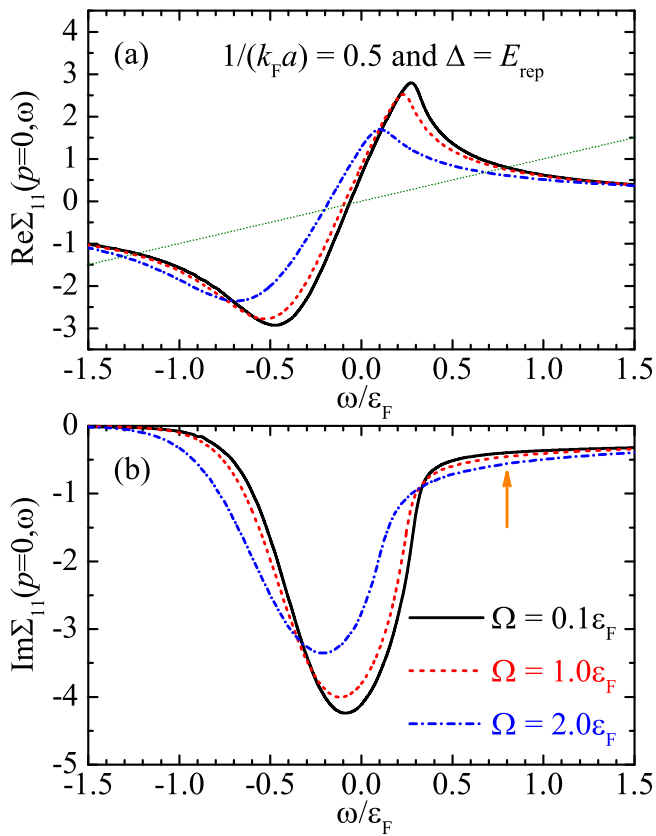}
\par\end{centering}
\caption{\label{fig2_seRabiVx05p} The real part (a) and imaginary part (b)
of the impurity self-energy $\Sigma_{11}(\mathbf{p},\omega)$ at zero
momentum ($\mathbf{p}=\mathbf{0}$) at the interaction strength $1/(k_{F}a)=0.5$.
The self-energy is in units of of $\varepsilon_{F}$. The temperature
is $T=0.2T_{F}$ and the detuning is equal to the repulsive polaron
energy (at zero Rabi coupling), $\Delta=E_{\textrm{rep}}$. We have
consider three characteristic Rabi couplings $\Omega=0.1\varepsilon_{F}$
(black solid line), $1.0\varepsilon_{F}$ (red dashed line), and $2.0\varepsilon_{F}$
(blue dot-dashed line). The green dotted line in (a) shows the curve
$y=\omega$. The arrow in (b) points to the repulsive polaron energy
$E_{\textrm{rep}}\simeq0.80\varepsilon_{F}$.}
\end{figure}

\subsection{Self-energy}

For the case of $a_{2}=0$, it is useful to contrast our $T$-matrix
result Eq. (\ref{eq:impurityGF}) with another approximated impurity
Green function \cite{DarkwahOppong2019,Adlong2021},
\begin{equation}
\mathbf{G}(\mathbf{p},\omega)=\left[\begin{array}{cc}
\omega-\epsilon_{\mathbf{p}}^{(I)}-\Sigma^{(0)}\left(\mathbf{p},\omega\right) & -\Omega/2\\
-\Omega/2 & \omega-\epsilon_{\mathbf{p}}^{(I)}-\Delta
\end{array}\right]^{-1},\label{eq:impurityGF_rabi0}
\end{equation}
where $\Sigma^{(0)}(\mathbf{p},\omega)$ is the impurity self-energy
of the spin-up state, determined in the \emph{absence} of the Rabi
coupling. Thus, the impurity Green function $G(\mathbf{p},\omega)=1/[\omega-\epsilon_{\mathbf{p}}^{(I)}-\Sigma^{(0)}(\mathbf{p},\omega)]$
describes a Fermi polaron when the impurity is \emph{always} kept
to the spin-up state. Earlier pioneering investigations of the dissipative
Fermi polaron Rabi dynamics \cite{Parish2016,Adlong2021} rely on
the applicability of Eq. (\ref{eq:impurityGF_rabi0}), which is justified
in the limit of small Rabi couplings (i.e., $\Omega\rightarrow0$),
where the correction arising from the Rabi coupling to the self-energy
$\Sigma^{(0)}(\mathbf{p},\omega)$ would scale like $(\Omega/\varepsilon_{F})^{2}$.
However, in the experiments \cite{Kohstall2012,Scazza2017,DarkwahOppong2019},
typically a reasonably large Rabi coupling $\Omega\sim0.7\varepsilon_{F}$
has to be taken, in order to have measurable signals. The validity
of Eq. (\ref{eq:impurityGF_rabi0}) for these Rabi coupling strengths
(i.e., $\Omega\sim\varepsilon_{F}$) then should be carefully examined.

The $T$-matrix result Eq. (\ref{eq:impurityGF}) is useful to check
such a validity, as the self-energy $\Sigma_{11}(\mathbf{p},\omega)$
is obtained in the presence of the Rabi coupling. In Fig. \ref{fig1_seRabiUnitary}
and Fig. \ref{fig2_seRabiVx05p}, we show the zero-momentum impurity
self-energy $\Sigma_{11}(\mathbf{p=0},\omega)$ at different Rabi
coupling strengths, in the unitary limit ($a=\pm\infty$) and at the
interaction strength $1/(k_{F}a)=0.5$, respectively. For these two
cases, we focus on the attractive and repulsive polaron branches,
respectively, by choosing the detuning $\Delta=E_{\textrm{att}}$
and $\Delta=E_{\textrm{rep}}$, where the energies of both attractive
($E_{\textrm{att}}$) and repulsive polarons ($E_{\textrm{rep}}$)
are determined without the Rabi coupling \cite{Hu2022}.

For the very small Rabi coupling $\Omega=0.1\varepsilon_{F}$, the
self-energy $\Sigma_{11}(\mathbf{p},\omega)$ is essentially $\Sigma^{(0)}(\mathbf{p},\omega)$.
By increasing $\Omega$ to $\varepsilon_{F}$, we can see a quantitative
modification to the self-energy. Although this modification is noticeable,
it still seems to be small enough to validate the approximate impurity
Green function in Eq. (\ref{eq:impurityGF_rabi0}). At the large Rabi
coupling $\Omega=2\varepsilon_{F}$, there are qualitative changes
to the self-energy, as indicated by the large shifts in the local
minimum and/or maximum positions in both $\textrm{Re}\Sigma_{11}(\mathbf{0},\omega)$
and $\textrm{Im}\Sigma_{11}(\mathbf{0},\omega)$. 

These changes indicate that, under the strong driving condition with
$\Omega\gg\varepsilon_{F}$ the quasiparticle properties of Fermi
polarons are strongly modified. Therefore, in the Rabi oscillation
experiments, we can no longer probe the polaron physics without Rabi
coupling, which is determined by the self-energy $\Sigma^{(0)}(\mathbf{p},\omega)$
at $\Omega=0$. In more detail, through strongly driven Rabi oscillations
we would instead measure $\Omega$-dependent attractive and repulsive
polaron energies, and their $\Omega$-dependent decay rates. The decay
rate is roughly proportional to $-\textrm{Im}\Sigma_{11}(\mathbf{0},\omega)$.
As indicated by the arrows in Fig. \ref{fig1_seRabiUnitary}(b) and
Fig. \ref{fig2_seRabiVx05p}(b), we find that the imaginary part of
the self-energy $-\textrm{Im}\Sigma_{11}(\mathbf{0},\omega)$ increases
with increasing Rabi coupling strength. The change in $-\textrm{Im}\Sigma_{11}(\mathbf{0},\omega\sim E_{\textrm{att}})$
of the attractive polaron branch is particularly significant at large
Rabi coupling: it increases from a negligible value $0.006\varepsilon_{F}$
to a considerable value $0.097\varepsilon_{F}$. As we shall see,
this will bring an additional damping to Fermi-polaron Rabi oscillations.

\begin{figure}
\begin{centering}
\includegraphics[clip,width=0.48\textwidth]{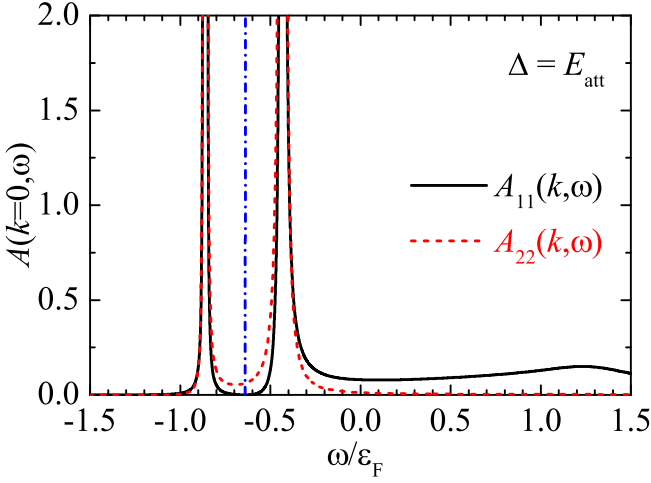}
\par\end{centering}
\caption{\label{fig3_akwRabi05Unitary} The zero-momentum impurity spectral
functions $A_{11}(\mathbf{p}=\mathbf{0},\omega)$ and $A_{22}(\mathbf{p}=\mathbf{0},\omega)$
in the unitary limit $1/a=0$ at the detunings $\Delta=E_{\textrm{att}}$.
The impurity spectral function is in units of of $\varepsilon_{F}^{-1}$.
The temperature is $T=0.2T_{F}$ and the Rabi coupling is $\Omega=0.5\varepsilon_{F}$.
The blue dot-dashed line indicates the peak position $\omega_{p}=E_{\textrm{att}}\simeq-0.64\varepsilon_{F}$
of the impurity spectral function without Rabi coupling $\Omega=0$.}
\end{figure}

\begin{figure}[t]
\begin{centering}
\includegraphics[width=0.48\textwidth]{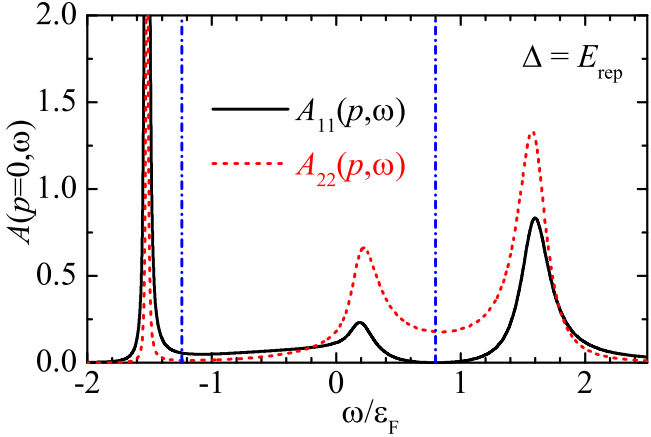}
\par\end{centering}
\caption{\label{fig4_akwRabi20Vx05p} The zero-momentum impurity spectral functions
$A_{11}(\mathbf{p}=\mathbf{0},\omega)$ and $A_{22}(\mathbf{p}=\mathbf{0},\omega)$
at the interaction strength $1/(k_{F}a)=0.5$ at the detunings $\Delta=E_{\textrm{rep}}$.
The impurity spectral function is in units of of $\varepsilon_{F}^{-1}$.
The temperature is $T=0.2T_{F}$ and the Rabi coupling is $\Omega=2.0\varepsilon_{F}$.
The blue dot-dashed line on the right indicates the peak position
$\omega_{p}=E_{\textrm{rep}}\simeq+0.80\varepsilon_{F}$ of the impurity
spectral functions $A_{11}$ and $A_{22}$ without Rabi coupling $\Omega=0$.
We note that, the impurity spectral functions $A_{11}$ has an additional
peak (of the attractive polaron) at $\omega_{p}=E_{\textrm{att}}\simeq-1.24\varepsilon_{F}$,
as indicated by the blue dot-dashed line on the left.}
\end{figure}

\subsection{Single-particle spectral function }

Using the self-energy $\Sigma_{11}(\mathbf{p},\omega)$ in Eq. (\ref{eq:impurityGF}),
we calculate directly the impurity Green functions $G_{11}(\mathbf{p},\omega)$
and $G_{22}(\mathbf{p},\omega)$, and the associated single-particle
spectral functions $A_{11}(\mathbf{p},\omega)$ and $A_{22}(\mathbf{p},\omega)$.
Two example cases of zero-momentum spectral function are shown in
Fig. \ref{fig3_akwRabi05Unitary} and Fig. \ref{fig4_akwRabi20Vx05p},
for the interaction strengths $1/(k_{F}a)=0$ and $1/(k_{F}a)=0.5$,
respectively. In each case, we consider the resonant detuning for
attractive or repulsive polarons.

\begin{figure}[t]
\begin{centering}
\includegraphics[width=0.45\textwidth]{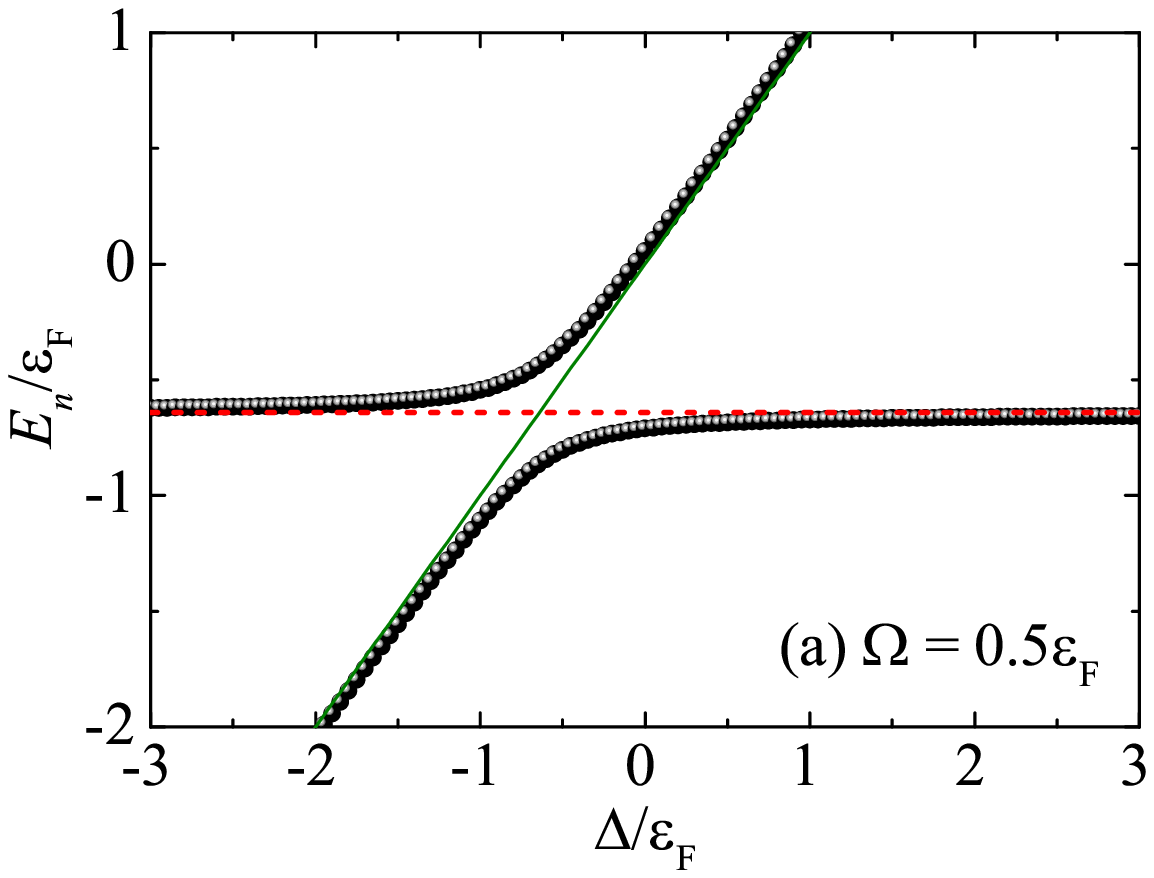}
\par\end{centering}
\begin{centering}
\includegraphics[width=0.45\textwidth]{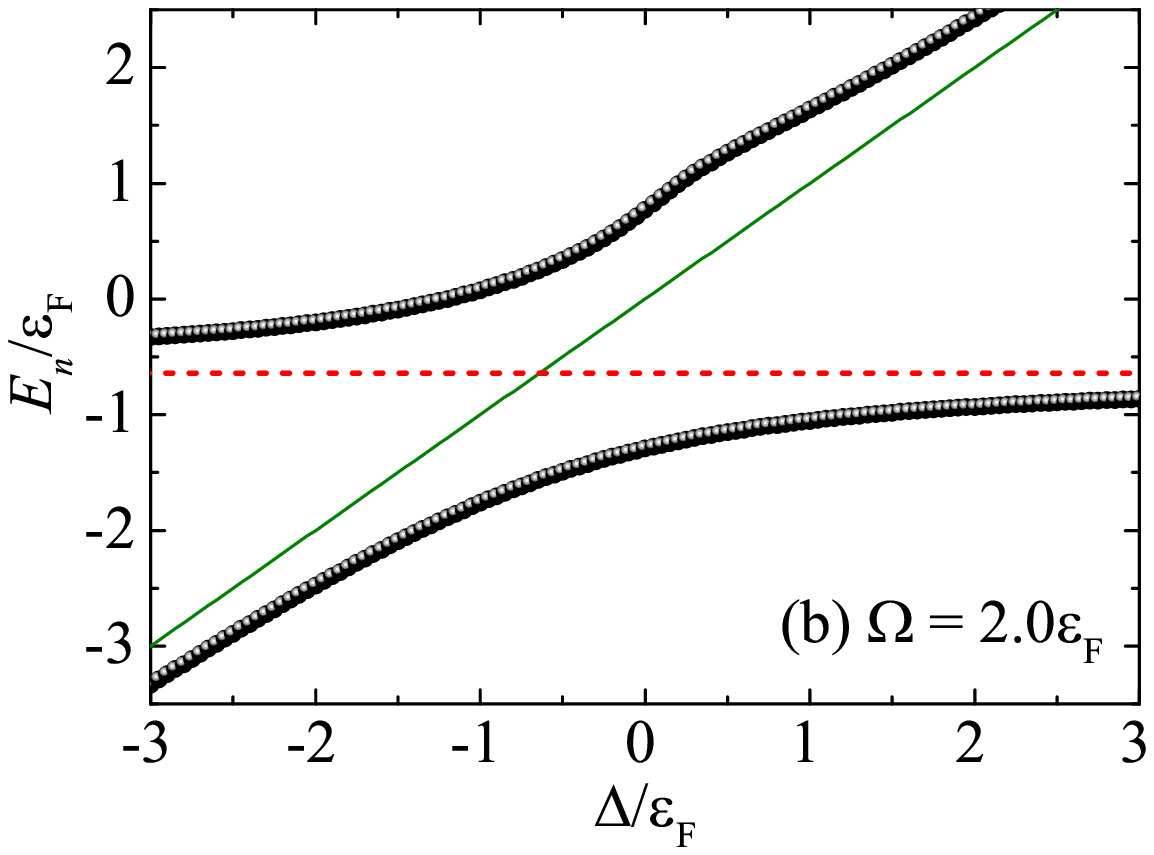}
\par\end{centering}
\caption{\label{fig5_energyUnitary} The zero-momentum energies of Fermi spin
polarons as a function of the detuning $\Delta$ in the unitary limit
$1/a$, at the Rabi frequencies $\Omega=0.5\varepsilon_{F}$ (a) and
$\Omega=2.0\varepsilon_{F}$ (b). The temperature is $T=0.2T_{F}$.
The green solid lines show $y=\Delta$, while the horizontal red dashed
lines indicate the attractive polaron energy (at zero Rabi coupling)
$E_{\textrm{att}}\simeq-0.64\varepsilon_{F}$.}
\end{figure}

\begin{figure}
\begin{centering}
\includegraphics[width=0.45\textwidth]{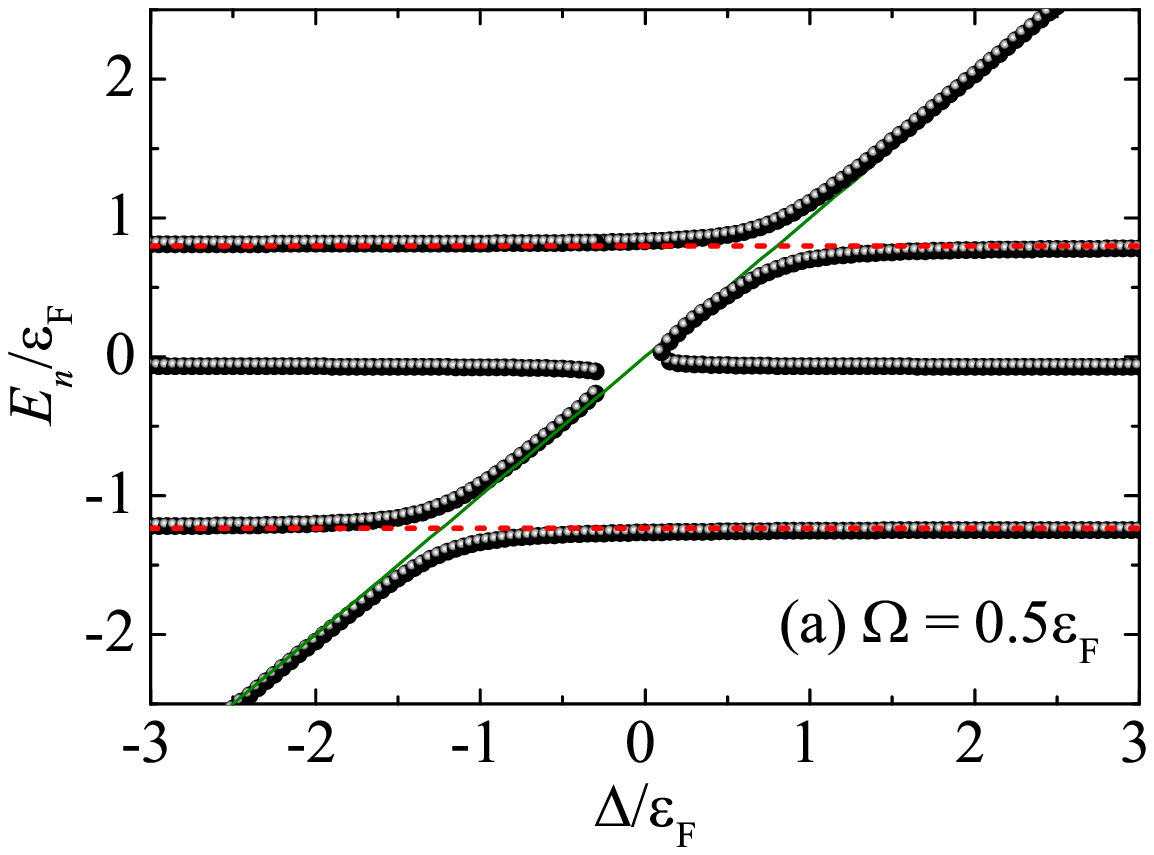}
\par\end{centering}
\begin{centering}
\includegraphics[width=0.45\textwidth]{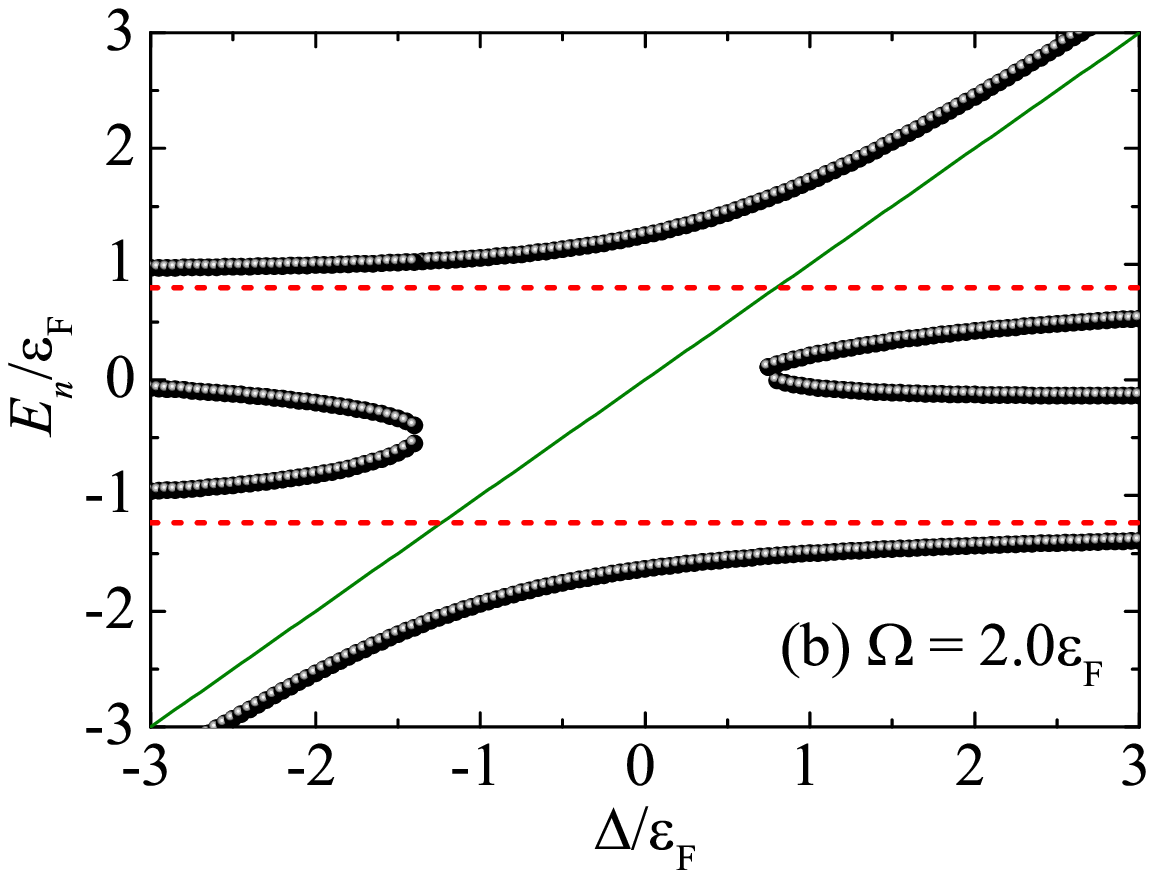}
\par\end{centering}
\caption{\label{fig6_energyVx05p} The zero-momentum energies of Fermi spin
polarons as a function of the detuning $\Delta$ at the interaction
strength $1/(k_{F}a)=0.5$ and at the Rabi frequencies $\Omega=0.5\varepsilon_{F}$
(a) and $\Omega=2.0\varepsilon_{F}$ (b). The temperature is $T=0.2T_{F}$.
The green solid lines show $y=\Delta$. The two horizontal red dashed
lines indicate the attractive polaron energy (at zero Rabi coupling)
$E_{\textrm{att}}\simeq-1.24\varepsilon_{F}$ and the repulsive polaron
energy (at zero Rabi coupling) $E_{\textrm{rep}}\simeq0.80\varepsilon_{F}$.}
\end{figure}

\begin{figure}
\begin{centering}
\includegraphics[width=0.48\textwidth]{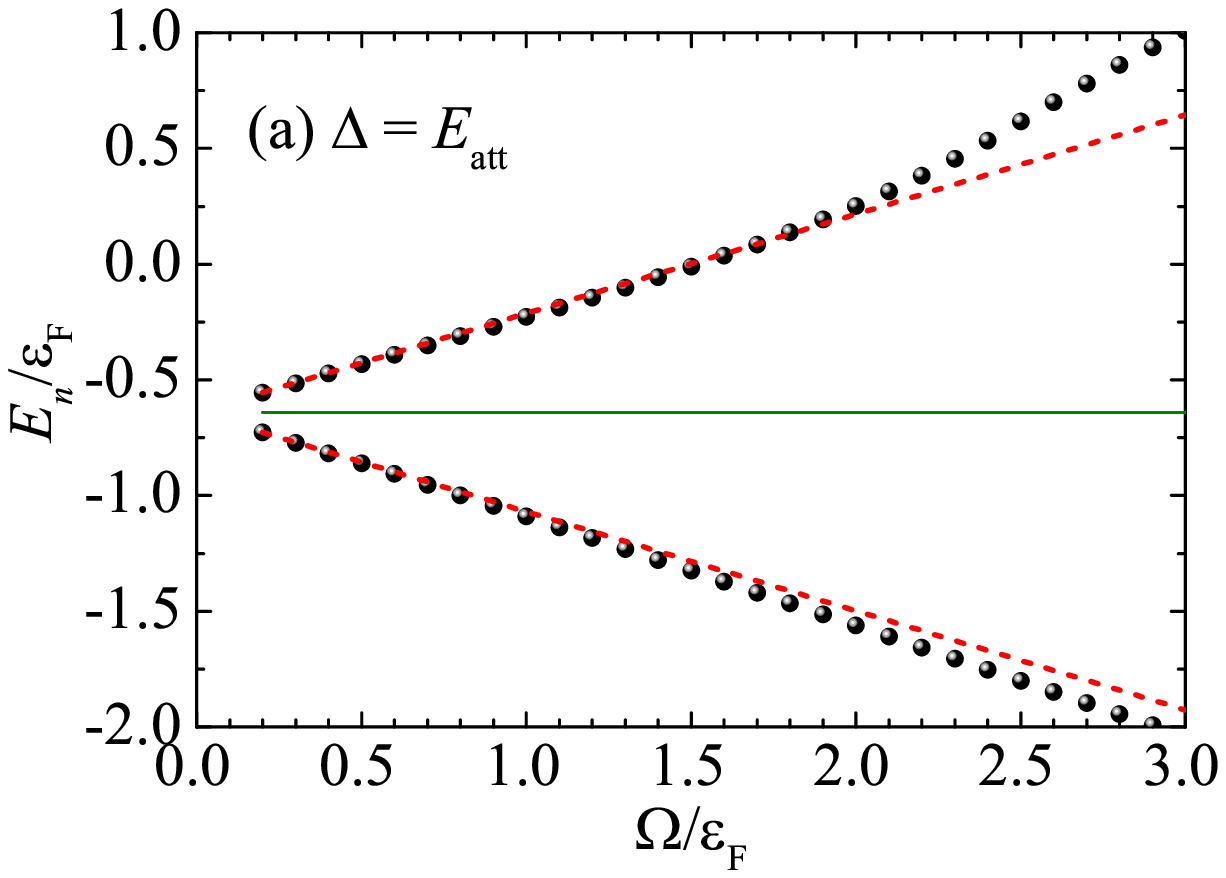}
\par\end{centering}
\begin{centering}
\includegraphics[width=0.48\textwidth]{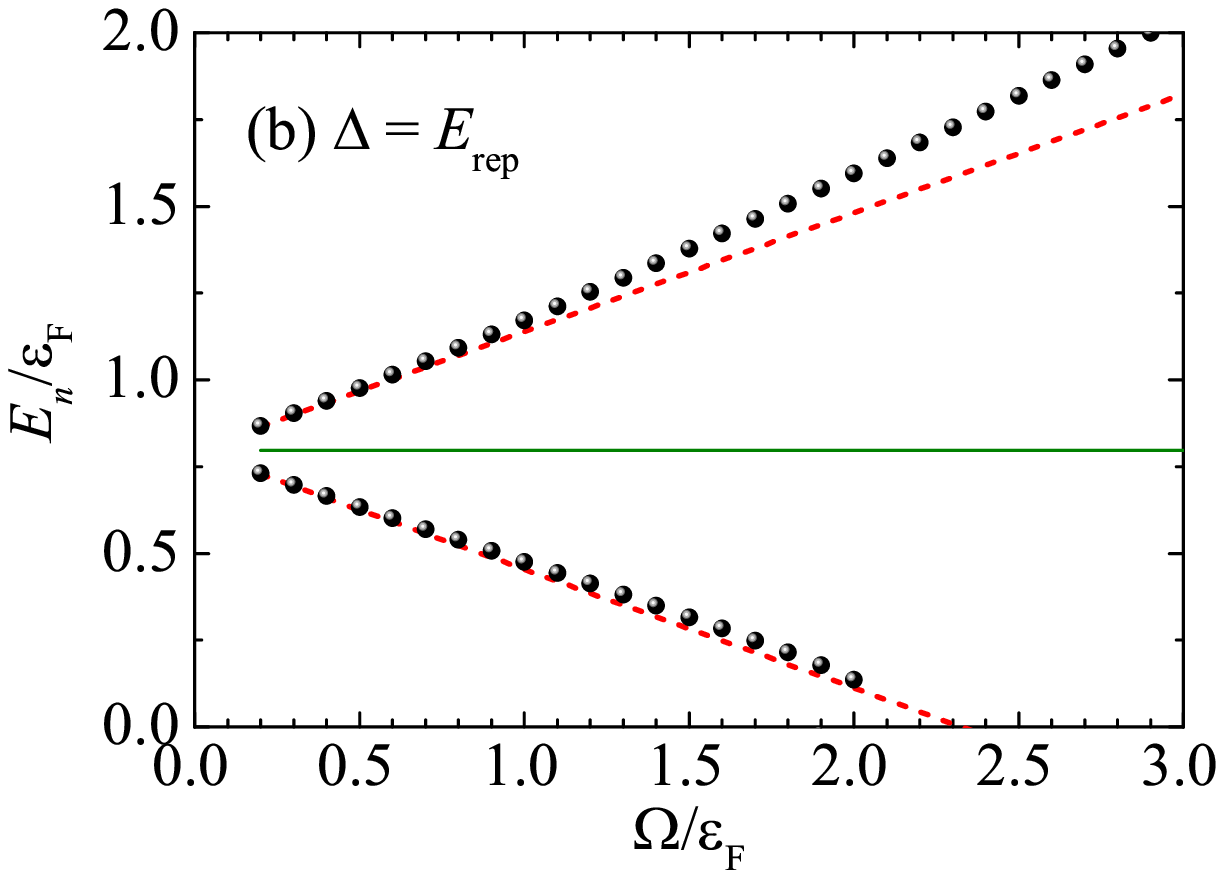}
\par\end{centering}
\caption{\label{fig7_RabiSplitting} The zero-momentum energies of Fermi spin
polarons as a function of the Rabi coupling $\Omega$ at the two interaction
strengths: $1/(k_{F}a)=0$ (a) and $1/(k_{F}a)=0.5$ (b). In the unitary
limit $1/(k_{F}a)=0$, we take the detuning $\Delta=E_{\textrm{att}}\simeq-0.64\varepsilon_{F}$,
while at the interaction strength $1/(k_{F}a)=0.5$, we set $\Delta=E_{\textrm{rep}}\simeq0.80\varepsilon_{F}$.
These two detunings are indicated by green solid lines. The red dashed
lines show the anticipated energies $E_{n}=\Delta\pm\sqrt{\mathcal{Z}}\Omega/2$,
where $\mathcal{Z}_{\textrm{att}}\simeq0.73$ in (a) and $\mathcal{Z}_{\textrm{rep}}\simeq0.47$
in (b). The temperature is $T=0.2T_{F}$.}
\end{figure}

It is readily seen that there are several peaks in the spectral functions
$A_{11}(\mathbf{0},\omega)$ and $A_{22}(\mathbf{0},\omega)$. Each
peak corresponds to a well-define quasiparticle, with its lifetime
characterized by the width of the peak. To determine the peak positions
or the quasiparticle energies, it is useful to approximate the impurity
Green function in Eq. (\ref{eq:impurityGF}) at zero momentum $\mathbf{p}=0$
as \cite{DarkwahOppong2019},
\begin{equation}
\mathbf{G}(\mathbf{0},\omega)=\left[\begin{array}{cc}
\mathcal{Z}^{-1}\left(\omega-\mathcal{E}_{P}+i\Gamma/2\right) & -\Omega/2\\
-\Omega/2 & \omega-\Delta
\end{array}\right]^{-1}.\label{eq:ImpurityGFApproximate}
\end{equation}
Here, we have assumed that the Green function $[\omega-\epsilon_{\mathbf{p}}^{(I)}-\Sigma_{11}(\mathbf{p},\omega)]^{-1}$
describes a Fermi polaron in the spin-up state with zero-momentum
polaron energies $\mathcal{E}_{P}$ satisfying $\mathcal{E}_{P}=\textrm{Re}\Sigma_{11}(\mathbf{0},\omega=\mathcal{E}_{P})$,
and therefore have approximated,
\begin{equation}
\frac{1}{\omega-\Sigma_{11}\left(\mathbf{0},\omega\right)}\simeq\frac{1}{\left(\omega-\mathcal{E}_{P}\right)\left(1-\frac{\partial\textrm{Re}\Sigma_{11}}{\partial\omega}\right)-i\textrm{Im\ensuremath{\Sigma_{11}}}},
\end{equation}
by Taylor-expanding $\Sigma_{11}(\mathbf{0},\omega)$ near $\omega=\mathcal{E}_{P}$.
Following the standard way \cite{Combescot2007,Hu2022} to introduce
the residue $\mathcal{Z}=[1-\partial\textrm{Re}\Sigma_{11}/\partial\omega]^{-1}$
and decay rate $\Gamma=-2\mathcal{Z\textrm{Im}}\Sigma_{11}$, we then
arrive at Eq. (\ref{eq:ImpurityGFApproximate}). This approximate
form of the impurity Green function is very useful to understand the
Rabi dynamics of the Fermi polaron in the spin-up state, as suggested
in Ref. \cite{DarkwahOppong2019}. For example, at the resonant detuning
$\Delta=\mathcal{E}_{P}$, it is easy to find that the poles of Eq.
(\ref{eq:ImpurityGFApproximate}) satisfy the equation,
\begin{equation}
\left(E-\mathcal{E}_{P}+i\frac{\Gamma}{2}\right)\left(E-\mathcal{E}_{P}\right)-\mathcal{Z}\frac{\Omega^{2}}{4}=0,
\end{equation}
and are given by ($\Gamma_{R}\equiv\Gamma/2$) \cite{DarkwahOppong2019,Adlong2020},
\begin{equation}
E_{\pm}=\left(\mathcal{E}_{P}\pm\frac{1}{2}\sqrt{\mathcal{Z}\Omega^{2}-\Gamma_{R}^{2}}\right)-i\frac{\Gamma_{R}}{2}.\label{eq:EnergyRabi}
\end{equation}
The form of the quasiparticle energies in the above equation clearly
indicates Rabi oscillations with a modified effective Rabi coupling
strength $\Omega_{\textrm{eff}}=\sqrt{\mathcal{Z}}\Omega$ and with
a damping rate $\Gamma_{R}=\Gamma/2$. Therefore, if one neglects
the $\Omega$-dependence of the residue $\mathcal{Z}$ and of the
decay rate $\Gamma$ (which seems justified from Fig. \ref{fig1_seRabiUnitary}
and Fig. \ref{fig2_seRabiVx05p} for $\Omega<\varepsilon_{F}$, as
we have discussed in the last subsection), one can directly extract
both residue and decay rate of Fermi polarons from Rabi oscillations
\cite{Kohstall2012,Scazza2017,DarkwahOppong2019}. In the strong driving
regime, $\Omega\gg\varepsilon_{F}$, instead we anticipate that the
effective Rabi coupling strength $\Omega_{\textrm{eff}}$ will deviate
from $\sqrt{\mathcal{Z}}\Omega$. More discussions on this point will
be provided in the next section.

As shown in Fig. \ref{fig3_akwRabi05Unitary}, for the unitary impurity-bath
interaction, where only attractive polaron exists, we find two peaks
with position well described by Eq. (\ref{eq:EnergyRabi}). The situation
becomes a little complicated at the interaction strength $1/(k_{F}a)=0.5$,
as reported in Fig. \ref{fig4_akwRabi20Vx05p}. There, we have both
attractive polaron and repulsive polaron in the spin-up state. As
we choose a resonant detuning $\Delta=\mathcal{E}_{P}=E_{\textrm{rep}}$
for the repulsive branch, there are two peaks locating at the positive
energy and roughly satisfying Eq. (\ref{eq:EnergyRabi}). However,
there is also a peak at about the attractive polaron energy $\mathcal{E}_{P}=E_{\textrm{att}}\sim-1.24\varepsilon_{F}$.
This peak is not taken into account in the approximate impurity Green
function Eq. (\ref{eq:ImpurityGFApproximate}), but it could be obtained
by numerically solving the poles of the full impurity Green function
Eq. (\ref{eq:impurityGF}). 

\subsection{Energies of Fermi spin polarons}

We have numerically determined the poles of the full impurity Green
function in Eq. (\ref{eq:impurityGF}), by neglecting the imaginary
part of the self-energy $\textrm{Im}\Sigma_{11}$. The results in
the unitary limit and at $1/(k_{F}a)=0.5$ as a function of the detuning
are shown in Fig. \ref{fig5_energyUnitary} and Fig. \ref{fig6_energyVx05p},
respectively, at two Rabi coupling strengths $\Omega=0.5\varepsilon_{F}$
(see the upper panels of the figures) and $\Omega=2\varepsilon_{F}$
(lower panels). 

In the unitary limit (Fig. \ref{fig5_energyUnitary}), we find two
energies of the Fermi spin polaron, which basically follow,
\begin{equation}
E_{\pm}=\frac{1}{2}\left[\mathcal{E}_{P}+\Delta\pm\sqrt{\mathcal{Z}\Omega^{2}-\left(\mathcal{E}_{P}-\Delta\right)^{2}}\right],\label{eq:EnergyRabiOffResonance}
\end{equation}
if we neglect the small decay rate (i.e., $\Gamma\ll\varepsilon_{F}$).
This is particularly evident at the small Rabi coupling (see Fig.
\ref{fig5_energyUnitary}(a)), where both energies follow $\mathcal{E}_{P}$
and $\Delta$ far off the resonance and exhibit a well-defined avoided
crossing at the resonance $\Delta=\mathcal{E}_{P}$. At the large
Rabi coupling (Fig. \ref{fig5_energyUnitary}(b)), however, we find
that the upper branch of the energies seems to develop an additional
structure around zero detuning $\Delta=0$. We attribute it to the
strongly modified self-energy $\Sigma_{11}$ at large Rabi coupling.

At the interaction strength $1/(k_{F}a)=0.5$ (Fig. \ref{fig6_energyVx05p}),
we typically find four poles in the impurity Green function Eq. (\ref{eq:impurityGF}).
The pole closest to $E=0$ is a \emph{false} solution, since we do
not include $\textrm{Im}\Sigma_{11}$ in finding the poles of Eq.
(\ref{eq:impurityGF}). In general, $\textrm{Im}\Sigma_{11}$ takes
a very large value near zero energy (see, i.e., Fig. \ref{fig2_seRabiVx05p}(b)).
Thus, it is meaningless to treat the near-zero-energy solution as
a well-defined quasiparticle. We find two avoided crossings located
at the resonances $\Delta=E_{\textrm{att}}$ and $\Delta=E_{\textrm{rep}}$.
Far off the resonances, the other three poles basically follow the
trace of $\mathcal{E}_{P}=E_{\textrm{att}}$, $\mathcal{E}_{P}=E_{\textrm{rep}}$
(see the red dotted lines) and $\Delta$ (thin green line). At large
Rabi coupling, the middle pole may disappear at the detuning $\Delta\sim0$,
as shown in Fig. \ref{fig6_energyVx05p}(b). This is again due to
the large value of $\textrm{Im}\Sigma_{11}$ near zero energy.

In Fig. \ref{fig7_RabiSplitting}, we report the energy splitting
at the resonant detuning, as a function of the Rabi coupling, as predicted
by Eq. (\ref{eq:EnergyRabi}). By ignoring the small decay rate $\Gamma_{R}$,
the energy splitting is given by $\delta E=\sqrt{\mathcal{Z}}\Omega$,
where $\mathcal{Z}$ is the residue of either attractive polaron or
repulsive polaron. For the Rabi coupling $\Omega\leq\varepsilon_{F}$,
we find that Eq. (\ref{eq:EnergyRabi}) provides an excellent fit
to the numerically extracted quasiparticle energies. At larger Rabi
coupling, nonlinear deviation from Eq. (\ref{eq:EnergyRabi}) becomes
sizable, indicating the breakdown of the approximate impurity Green
function in Eq. (\ref{eq:ImpurityGFApproximate}).

\section{Dissipative Rabi dynamics of Fermi polarons}

Let us now try to better understand the recent experiments on Fermi
polaron Rabi oscillations \cite{Kohstall2012,Scazza2017,DarkwahOppong2019},
by examining more closely the role play by a reasonably large Rabi
coupling strength $\Omega\sim\varepsilon_{F}$. Experimentally, the
impurity is initially prepared in the non-interacting (or weakly-interacting)
spin-down state. At time zero, a simple square pulse is added to transfer
the impurity to the spin-up state that is in strongly interaction
with the Fermi bath. The frequency of the pulse is suitably chosen,
so either the attractive polaron branch or the repulsive polaron branch
of the spin-up state is selected to be on resonance (i.e., $\Delta=E_{\textrm{att}}$
or $\Delta=E_{\textrm{rep}}$). After a variable holding time $t$,
the relative population of the impurity in the spin-down state is
then determined.

\subsection{Theory of the dissipative Rabi oscillation}

Physically, for a single impurity this procedure measures the spin-down
occupation,
\begin{align}
n_{\downarrow}\left(t\right) & =\left\langle \psi\left(t\right)\left|\sum_{\mathbf{p}}d_{\mathbf{p}\downarrow}^{\dagger}d_{\mathbf{p}\downarrow}\right|\psi\left(t\right)\right\rangle ,\nonumber \\
 & =\left\langle \psi\left(0\right)\left|\sum_{\mathbf{p}}d_{\mathbf{p}\downarrow}^{\dagger}\left(t\right)d_{\mathbf{p}\downarrow}\left(t\right)\right|\psi\left(0\right)\right\rangle \label{eq:nt0}
\end{align}
where the time-dependent many-body wavefunction is $\left|\psi\left(t\right)\right\rangle =e^{-i\mathcal{H}t/\hbar}\left|\psi\left(0\right)\right\rangle $,
time-dependent field operator $d_{\mathbf{p}\downarrow}(t)=e^{i\mathcal{H}t/\hbar}d_{\mathbf{p}\downarrow}e^{-i\mathcal{H}t/\hbar}$,
and the initial wavefunction at time $t=0$ is given by, 
\begin{equation}
\left|\psi\left(0\right)\right\rangle =\left|\downarrow\right\rangle _{T}\otimes\left|\textrm{FS}\right\rangle .\label{eq:fai0}
\end{equation}
Here, since the initial spin-down impurity does not interact with
the thermal Fermi bath, we have taken $\left|\psi\left(0\right)\right\rangle $
as a \emph{direct} product of a \emph{thermal} impurity state $\left|\downarrow\right\rangle _{T}$
and a thermal Fermi sea $\left|\textrm{FS}\right\rangle $. In the
Fermi sea, at finite temperature $T$ fermionic atoms are occupied
into single-particle states according to the Fermi-Dirac distribution.
For a single impurity, the thermal probability of the spin-down impurity
would be given by a suitable distribution function $f_{\mathbf{k}}$
that is determined by the type or statistics of the impurity, if it
has a momentum $\mathbf{k}$. By substituting Eq. (\ref{eq:fai0})
into Eq. (\ref{eq:nt0}), we find that,
\begin{equation}
n_{\downarrow}\left(t\right)=\sum_{\mathbf{p}\mathbf{k}}f_{\mathbf{k}}\left\langle \textrm{FS}\left|d_{\mathbf{k}\downarrow}d_{\mathbf{p}\downarrow}^{\dagger}\left(t\right)d_{\mathbf{p}\downarrow}\left(t\right)d_{\mathbf{k}\downarrow}^{\dagger}\right|\textrm{FS}\right\rangle .\label{eq:nt4f}
\end{equation}

It is difficult to exactly evaluate $n_{\downarrow}(t)$ for long
time, which involves a product of four field operators. This is because
the effects due to strong correlations will gradually accumulate during
the time evolution. For the time scale in the Rabi dynamics experiments
(i.e., for a few Rabi oscillations), however, it might be instructive
to consider the following first-order, \emph{mean-field} type decoupling,
\begin{align}
n_{\downarrow}\left(t\right) & \simeq\sum_{\mathbf{p}}f_{\mathbf{p}}\left\langle d_{\mathbf{p}\downarrow}d_{\mathbf{p}\downarrow}^{\dagger}\left(t\right)\right\rangle \left\langle d_{\mathbf{p}\downarrow}\left(t\right)d_{\mathbf{p}\downarrow}^{\dagger}\right\rangle ,\nonumber \\
 & =\sum_{\mathbf{p}}f_{\mathbf{p}}\mathcal{S}_{\downarrow\downarrow}\left(\mathbf{p},-t\right)\mathbf{\mathcal{S}}_{\downarrow\downarrow}\left(\mathbf{p},t\right),
\end{align}
as inspired by the well-known Wick theorem in the diagrammatic theory.
Here, in the second line we have introduced for $t>0$,
\begin{equation}
\mathbf{\mathcal{S}_{\downarrow\downarrow}}\left(\mathbf{p},t\right)\equiv\left\langle \textrm{FS}\left|d_{\mathbf{p}\downarrow}\left(t\right)d_{\mathbf{p}\downarrow}^{\dagger}\right|\textrm{FS}\right\rangle \equiv\left\langle d_{\mathbf{p}\downarrow}\left(t\right)d_{\mathbf{p}\downarrow}^{\dagger}\right\rangle .
\end{equation}
It is easy to recognize that $\mathbf{\mathcal{S}_{\downarrow\downarrow}}(\mathbf{p},t)=iG_{22}(\mathbf{p},t)$
is exactly the impurity Green function in the spin-down channel in
the time domain. Therefore, we can determine it directly from the
single-particle spectral functions, i.e., 
\begin{equation}
\mathbf{\mathcal{S}_{\downarrow\downarrow}}\left(\mathbf{p},t\right)=\intop_{-\infty}^{+\infty}d\omega A_{22}\left(\mathbf{p},\omega\right)e^{-i\omega t}.
\end{equation}
 and consequently, we are able to calculate 
\begin{equation}
n_{\downarrow}(t)\simeq\sum_{\mathbf{p}}f_{\mathbf{p}}\left|\mathbf{\mathcal{S}}_{\downarrow\downarrow}\left(\mathbf{p},t\right)\right|^{2}.\label{eq:ntdd}
\end{equation}
An expression similar to Eq. (\ref{eq:ntdd}) but without the thermal
average has been advised by Adlong \textit{et al.} (see Eq. (S46)
in Ref. \cite{Adlong2020}), based on the variational Chevy's ansatz.

To show the usefulness of Eq. (\ref{eq:ntdd}), let us derive an analytic
expression of $n_{\downarrow}(t)$ following Ref. \cite{Adlong2020},
by using the approximate impurity Green function in Eq. (\ref{eq:ImpurityGFApproximate})
at zero momentum. It is readily seen that, the spin-down impurity
Green function then takes the approximate form,
\begin{equation}
G_{22}\left(\mathbf{0},\omega\right)=\frac{A}{\omega-E_{+}+i\Gamma_{R}/2}+\frac{1-A}{\omega-E_{-}+i\Gamma_{R}/2},
\end{equation}
where the pole energies $E_{+}$ and $E_{-}$ are given by Eq. (\ref{eq:EnergyRabiOffResonance}),
and $A\equiv1/2-(\mathcal{E}_{P}-\Delta)/(2\Omega_{\textrm{eff}})$
with $\Omega_{\textrm{eff}}=\sqrt{\mathcal{Z}\Omega^{2}+(\mathcal{E}_{P}-\Delta)^{2}}$.
After some straightforward algebra, we find that,
\begin{equation}
\left|\mathbf{\mathcal{S}_{\downarrow\downarrow}}\right|^{2}\simeq e^{-\Gamma_{R}t}\left[\cos^{2}\frac{\Omega_{\textrm{eff}}t}{2}+\frac{\left(\mathcal{E}_{P}-\Delta\right)^{2}}{\Omega_{\textrm{eff}}^{2}}\sin^{2}\frac{\Omega_{\textrm{eff}}t}{2}\right],\label{eq:s2dd}
\end{equation}
which clearly exhibits an oscillation with periodicity $2\pi/\Omega_{\textrm{eff}}$
and damping rate $\Gamma_{R}$. 

The above equation and Eq. (\ref{eq:ntdd}) are not applicable for
long evolution time. This is partly reflected in the exponential decay
of $\left|\mathbf{\mathcal{S}_{\downarrow\downarrow}}\right|^{2}$,
which implies that $n_{\downarrow}(t\rightarrow\infty)=0$ for any
detuning $\Delta$. However, at the resonant detuning $\Delta=\mathcal{E}_{P}$,
the effective bias for the impurity spin would be zero \cite{Knap2013}.
Therefore, we should anticipate a zero steady-state magnetization,
or, $n_{\uparrow}(t\rightarrow\infty)=n_{\downarrow}(t\rightarrow\infty)=1/2$
\cite{Knap2013}. A possible reason why Eq. (\ref{eq:ntdd}) can not
give a zero steady-state magnetization is that the single impurity
condition, i.e., $\sum_{\mathbf{p}}[d_{\mathbf{p}\uparrow}^{\dagger}(t)d_{\mathbf{p}\uparrow}(t)+d_{\mathbf{p}\downarrow}^{\dagger}(t)d_{\mathbf{p}\downarrow}(t)]=1$,
is not strictly satisfied by our approximated mean-field type decoupling.

To rectify this weakness, it is useful to consider the following operator
for the spin-down occupation,
\begin{equation}
\hat{n}_{\downarrow}\left(t\right)=\frac{1}{2}-\frac{1}{2}\sum_{\mathbf{p}}\left[d_{\mathbf{p}\uparrow}^{\dagger}\left(t\right)d_{\mathbf{p}\uparrow}\left(t\right)-d_{\mathbf{p}\downarrow}^{\dagger}\left(t\right)d_{\mathbf{p}\downarrow}\left(t\right)\right],
\end{equation}
and then calculate 
\begin{equation}
n_{\downarrow}\left(t\right)=\sum_{\mathbf{k}}f_{\mathbf{k}}\left\langle \textrm{FS}\left|d_{\mathbf{k}\downarrow}\hat{n}_{\downarrow}\left(t\right)d_{\mathbf{k}\downarrow}^{\dagger}\right|\textrm{FS}\right\rangle .
\end{equation}
By using the mean-field decoupling and repeating the steps that lead
to Eq. (\ref{eq:ntdd}), it is easy to derive that,
\begin{equation}
n_{\downarrow}(t)\simeq\frac{1}{2}+\frac{1}{2}\sum_{\mathbf{p}}f_{\mathbf{p}}\left[\left|\mathbf{\mathcal{S}}_{\downarrow\downarrow}\left(\mathbf{p},t\right)\right|^{2}-\left|\mathbf{\mathcal{S}}_{\uparrow\downarrow}\left(\mathbf{p},t\right)\right|^{2}\right],\label{eq:ntnew}
\end{equation}
where $\mathbf{\mathcal{S}}_{\uparrow\downarrow}(\mathbf{p},t)$ takes
the form, 
\begin{equation}
\mathbf{\mathcal{S}_{\uparrow\downarrow}}\left(\mathbf{p},t\right)=\intop_{-\infty}^{+\infty}d\omega A_{12}\left(\mathbf{p},\omega\right)e^{-i\omega t}.
\end{equation}
The use of Eq. (\ref{eq:ntnew}) is still restricted to the short-time
evolution of a few Rabi oscillations. However, we anticipate that
it may provide a more accurate prediction than Eq. (\ref{eq:ntdd})
at the resonant detuning, the case that we will focus on.

\begin{figure}
\begin{centering}
\includegraphics[width=0.48\textwidth]{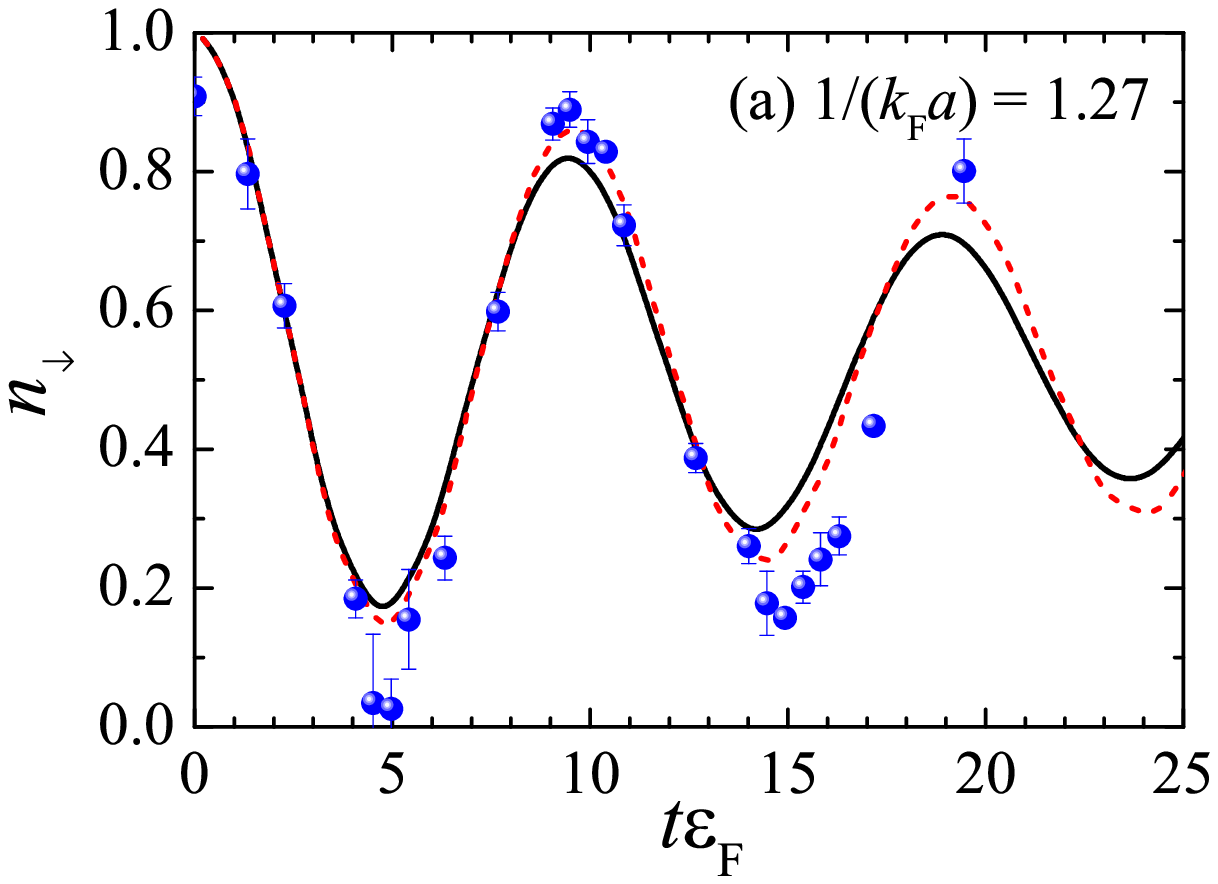}
\par\end{centering}
\begin{centering}
\includegraphics[width=0.48\textwidth]{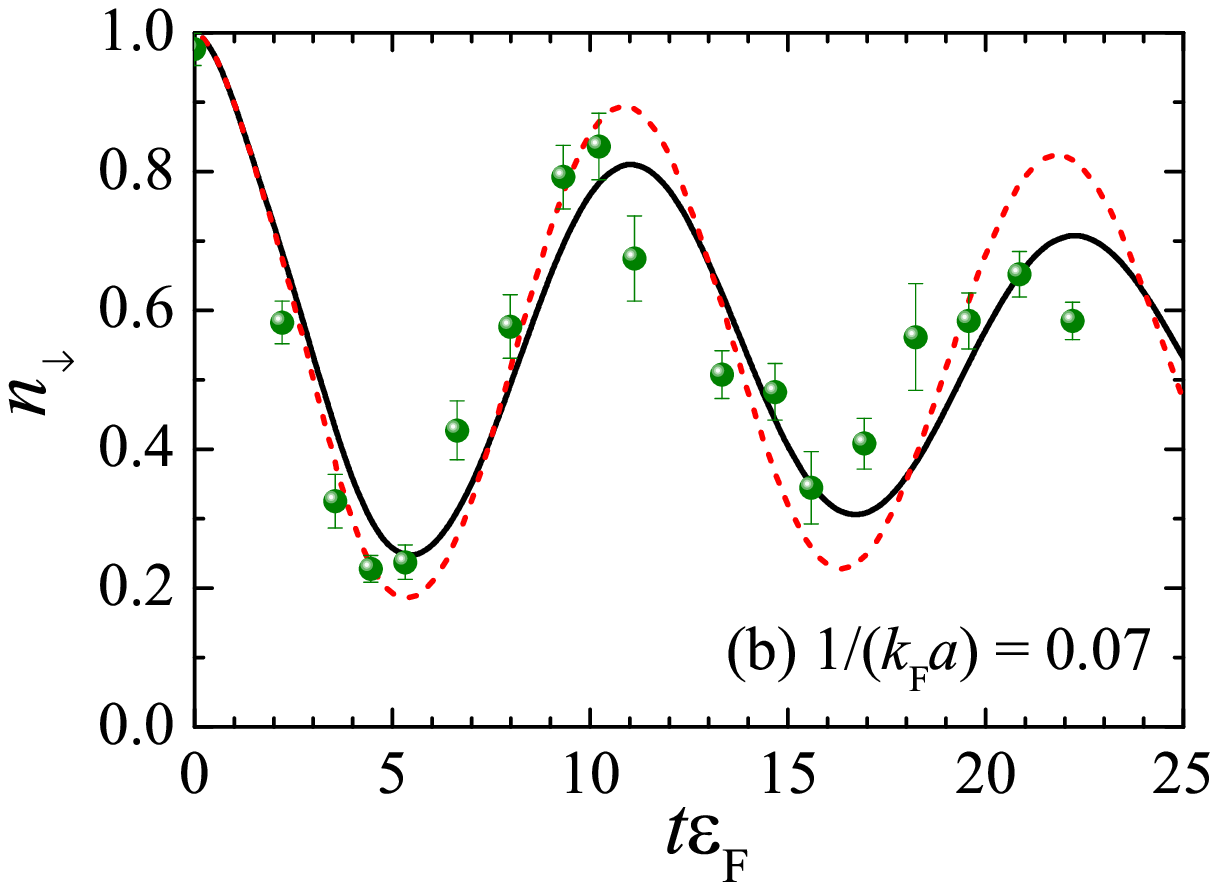}
\par\end{centering}
\caption{\label{fig8_expt} The comparison of the theory (lines) with the experimental
data from the LENS group (symbols) \cite{Ness2020}, for the Rabi
oscillations of a repulsive polaron at $1/(k_{F}a)=1.27$ (a) and
of an attractive polaron near the unitary limit $1/(k_{F}a)\simeq0$
(b). The data in (a) are extracted from Fig. 2(e) of Ref. \cite{Adlong2020}
and the data in (b) are extracted from Fig. S4(b) of Ref. \cite{Scazza2017}.
For the red dashed lines, we consider the Rabi oscillations of the
impurity with zero-momentum. For the black solid lines, we include
the momentum average, arising from the thermal distribution of the
momentum at finite temperature. In the theoretical calculations, we
always take the detuning $\Delta$ that is resonant with either the
repulsive polaron energy $E_{\textrm{rep}}$ or attractive polaron
energy $E_{\textrm{att}}$ at zero Rabi coupling. The impurity density
is taken as $n_{\textrm{imp}}/n=0.15$, the Rabi coupling is $\Omega=0.7\varepsilon_{F}$
and the temperature is $T=0.13T_{F}$, following the experimental
condition \cite{Scazza2017}. In the comparison, we do not include
any adjustable free parameters.}
\end{figure}

\subsection{Comparison between the theory and experiments}

We consider the recent Rabi oscillation experiment carried out at
the European Laboratory for Non-linear Spectroscopy (LENS) \cite{Scazza2017}.
There, impurities are the minority fermionic $^{6}$Li atoms, initially
in the weakly interacting hyperfine state $\left|2\right\rangle $
(i.e., the second lowest-energy Zeeman state). The impurity concentration
is about $n_{\textrm{imp}}\simeq0.15n$, where $n$ is the density
of the majority $^{6}$Li atoms in the hyperfine state $\left|1\right\rangle $.
The temperature is about $T\simeq0.13T_{F}$, where the Fermi energy
$T_{F}$ is determined by the density $n$. Therefore, in the Rabi
measurement, initially the impurities would follow a Fermi-Dirac distribution
of an ideal Fermi gas, $f_{\mathbf{p}}=f[\epsilon_{\mathbf{p}}^{(I)}-\mu_{I}]$,
where the impurity chemical potential $\mu_{I}$ can be determined
by solving the number equation,
\begin{equation}
\sum_{\mathbf{p}}f_{\mathbf{p}}=\sum_{\mathbf{p}}\frac{1}{\exp\left[\frac{\epsilon_{\mathbf{p}}^{(I)}-\mu_{I}}{k_{B}T}\right]+1}=n_{\textrm{imp}}.
\end{equation}
Theoretically, we have solved the impurity spectral functions $A_{12}(\mathbf{p},\omega)$
and $A_{22}(\mathbf{p},\omega)$ at the given experimental Rabi coupling
strength $\Omega\simeq0.7\varepsilon_{F}$ and at different interaction
parameters $1/(k_{F}a)$, and have consequently calculated $\mathbf{\mathcal{S}}_{\uparrow\downarrow}(\mathbf{p},t)$
and $\mathbf{\mathcal{S}}_{\downarrow\downarrow}(\mathbf{p},t)$.
By integrating over the momentum with the distribution function $f_{\mathbf{p}}$,
we then determine the time-dependence of the spin-down occupation
$n_{\downarrow}(t)$ in Eq. (\ref{eq:ntnew}).

In Fig. \ref{fig8_expt}, we compare our theoretical predictions (lines)
with the experimental data (circles) for the repulsive polaron at
$1/(k_{F}a)=1.27$ (a) and for the attractive polaron near the unitary
limit $1/(k_{F}a)=0.07$ (b) \cite{Scazza2017}. The red dashed line
indicates the result for the zero-momentum polaron, without taking
into account the thermal average over the momentum distribution, while
the black solid line includes the momentum average at finite temperature.
We find a good agreement between theory and experiment \cite{Scazza2017},
without any free adjustable parameters. In particular, for the attractive
polaron in Fig. \ref{fig8_expt}(b), most of the experimental data
locate on the solid line within the experimental error bar. The good
agreement partly justifies the approximated mean-field decoupling
used to derive Eq. (\ref{eq:ntnew}) for the short-time evolution
of $n_{\downarrow}(t)$. 

For the repulsive polaron, the agreement also justifies the experimental
procedure of extracting the residue of the polaron $\mathcal{Z}$
from the oscillation periodicity and the effective Rabi coupling strength
(i.e., $\mathcal{Z}\simeq\Omega_{\textrm{eff}}^{2}/\Omega^{2}$) and
of measuring the polaron decay rate $\Gamma$ from the damping of
Rabi oscillations (i.e., $\Gamma=2\Gamma_{R}$). However, it should
be emphasized that, strictly speaking, the obtained polaron residue
and decay rate are not for the zero-momentum polaron at nonzero Rabi
coupling, as assumed in the recent theoretical analysis \cite{Adlong2020}
(see, nevertheless, more discussions on the role played by the finite
Rabi coupling in Sec. IVD). They are contributed from polarons with
different momenta thermally distributed according to $f_{\mathbf{p}}$.
This is clearly evidenced by the difference between the dashed line
and solid line, as shown in Fig. \ref{fig8_expt}(a). Although the
difference due to finite momentum is small, it can lead to a \emph{quantitative
}modification to, for example, the theoretically predicted damping
for Rabi oscillations.

For the attractive polaron in Fig. \ref{fig8_expt}(b), the difference
between the dashed line and solid line is even larger. In this case,
it is worth noting that the damping rate of Rabi oscillations does
not correspond to the decay rate of Fermi polarons. Even at zero momentum,
the damping rate exhibited by the red dashed line is much larger than
the decay rate of the attractive polaron. The latter is actually \emph{negligible}
at $T=0.13T_{F}$ \cite{Hu2022}. This inequivalence comes from the
fact that the imaginary part of the self-energy $\textrm{Im}\Sigma_{11}$
changes \emph{dramatically} near the attractive polaron energy, as
indicated by the arrow in Fig. \ref{fig1_seRabiUnitary}(b). As a
result, although the Taylor expansion of $\textrm{Re}\Sigma_{11}$
is still meaningful, the expansion of $\textrm{Im}\Sigma_{11}$ near
the attractive polaron energy becomes problematic for large Rabi coupling.
The use of the approximate impurity Green function Eq. (\ref{eq:ImpurityGFApproximate})
then will strongly under-estimate the decay rate at the experimental
Rabi coupling strength $\Omega\simeq0.7\varepsilon_{F}$. In sharp
contrast, $\textrm{Im}\Sigma_{11}$ has a very weak energy-dependence
near the repulsive polaron energy, as seen from Fig. \ref{fig2_seRabiVx05p}(b).
The approximate impurity Green function Eq. (\ref{eq:ImpurityGFApproximate})
is an excellent approximation at $\Omega\simeq0.7\varepsilon_{F}$
for repulsive Fermi polarons.

\begin{figure}
\begin{centering}
\includegraphics[width=0.48\textwidth]{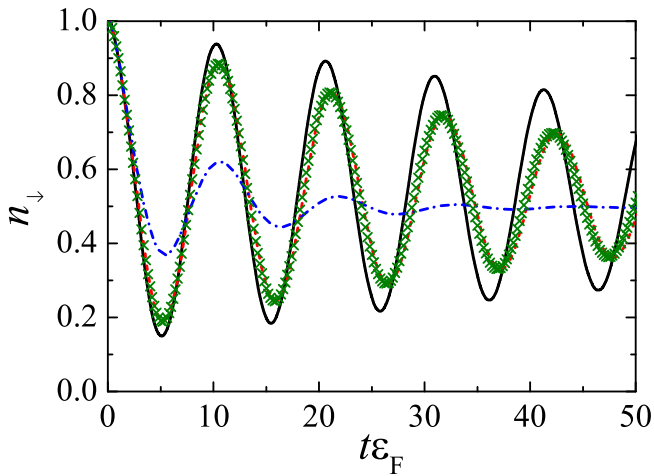}
\par\end{centering}
\caption{\label{fig9_kdepRabiOSCUnitary} Rabi oscillations of an attractive
polaron in the unitary limit $1/(k_{F}a)=0$ at different momenta:
$k=0$ (black solid line), $0.5k_{F}$ (red dashed line), and $k_{F}$
(blue dot-dashed line). The Rabi oscillation after the momentum average
is shown by symbols (green crosses). Here, the temperature is $T=0.1T_{F}$,
the detuning $\Delta=E_{\textrm{att}}\simeq-0.61\varepsilon_{F}$
and the Rabi coupling $\Omega=0.7\varepsilon_{F}$. For the momentum
average, we take the impurity density $n_{\textrm{imp}}/n=0.15$.}
\end{figure}

\subsection{Importance of the momentum average}

Let us now examine more carefully the effect of the momentum average
for Rabi oscillations of the attractive polaron. In Fig. \ref{fig9_kdepRabiOSCUnitary},
we shown the oscillations in $n_{\downarrow}(t)$ contributed from
the momentum $k=0$ (black solid line), $0.5k_{F}$ (red dashed line),
and $k_{F}$ (blue dot-dashed line). In comparison to the zero-momentum
oscillation, a finite momentum gradually increases the periodicity
of Rabi oscillations, in addition to causing more damping. In particular,
at large momentum (i.e., the $k=k_{F}$ curve), the oscillation becomes
overdamped. After taking into account the thermal distribution function
$f_{\mathbf{k}}$, the final theoretical prediction with momentum
average roughly follow the curve at $0.5k_{F}$ at the given low temperature
$T=0.1T_{F}$. 

\begin{figure}
\begin{centering}
\includegraphics[width=0.48\textwidth]{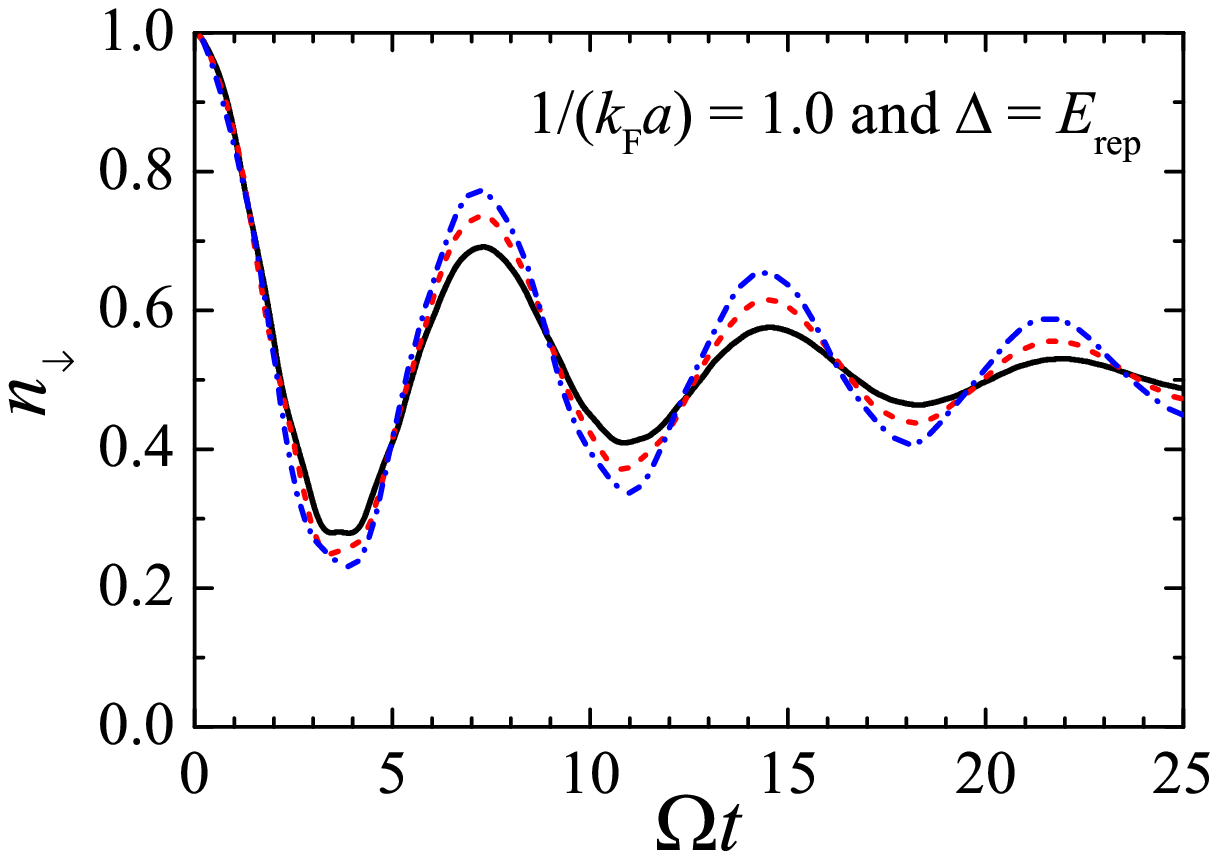}
\par\end{centering}
\caption{\label{fig10_vx10pRabiOSC} Momentum-averaged Rabi oscillations of
a repulsive polaron at the interaction strength $1/(k_{F}a)=1$ at
different Rabi couplings: $\Omega=0.5\varepsilon_{F}$ (black solid
line), $0.7\varepsilon_{F}$ (red dashed line), and $\varepsilon_{F}$
(blue dot-dashed line). Here, the temperature is $T=0.1T_{F}$ and
the detuning is $\Delta=E_{\textrm{rep}}\simeq0.53\varepsilon_{F}$.
For the momentum average, we take the impurity density $n_{\textrm{imp}}/n=0.15$.}
\end{figure}

\begin{figure}
\begin{centering}
\includegraphics[width=0.48\textwidth]{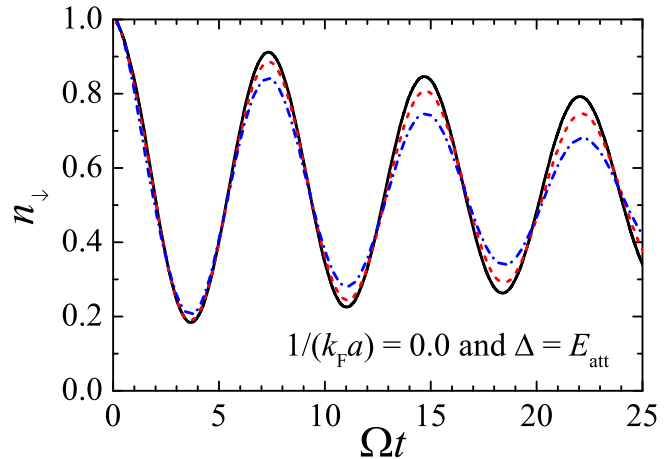}
\par\end{centering}
\caption{\label{fig11_UnitaryRabiOSC} Momentum-averaged Rabi oscillations
of an attractive polaron in the unitary limit $1/(k_{F}a)=0$ at different
Rabi couplings: $\Omega=0.5\varepsilon_{F}$ (black solid line), $0.7\varepsilon_{F}$
(red dashed line), and $\varepsilon_{F}$ (blue dot-dashed line).
Here, the temperature is $T=0.1T_{F}$ and the detuning is $\Delta=E_{\textrm{att}}\simeq-0.61\varepsilon_{F}$.
For the momentum average, we take the impurity density $n_{\textrm{imp}}/n=0.15$.}
\end{figure}

\subsection{Dependence on the Rabi coupling}

Here we examine the dependence of Rabi oscillations on the Rabi coupling
strength. In Fig. \ref{fig10_vx10pRabiOSC} and Fig. \ref{fig11_UnitaryRabiOSC},
we respectively report the Rabi oscillations of the repulsive and
attractive Fermi polarons at three Rabi coupling strengths: $\Omega=0.5\varepsilon_{F}$
(black solid line), $0.7\varepsilon_{F}$ (red dashed line), and $\varepsilon_{F}$
(blue dot-dashed line).

For the repulsive polaron at $1/(k_{F}a)=1$, the visibility or amplitude
of oscillations increases with increasing Rabi coupling, indicating
a smaller damping rate. This counterintuitive tendency cannot be simply
understood from the picture of a zero-momentum Fermi polaron, whose
$-\textrm{Im}\Sigma_{11}$ near the repulsive polaron energy would
increase with increasing Rabi coupling, as shown in Fig. \ref{fig2_seRabiVx05p}(b).
Therefore, the momentum average tends to \emph{decrease} the damping
rate of Rabi oscillations at large Rabi coupling. On the other hand,
a finite momentum increases the damping rate at a fixed Rabi coupling,
as we already discussed. It is then readily seen that the Rabi coupling
and finite momentum have opposite effects on the damping of Rabi oscillations.
Although these two effects may not completely cancel, it seems reasonable
to interpret the observed damping of Rabi oscillation (at finite Rabi
coupling with momentum average) as the decay rate of zero-momentum
Fermi polaron (at \emph{zero} Rabi coupling). This understanding therefore
supports the observation found in the LENS experiment \cite{Scazza2017}
that, the damping rate of Rabi oscillations quantitatively matches
the predicted quasiparticle peak spectral width $\Gamma$ of repulsive
Fermi polarons at zero momentum.

For the attractive polaron in the unitary limit, in contrast, the
visibility of Rabi oscillations decreases with increasing Rabi coupling.
This enhanced damping can be understood from the rapidly changing
$-\textrm{Im}\Sigma_{11}$ near the attractive polaron energy, as
shown in Fig. \ref{fig1_seRabiUnitary}(b). If we consider the approximated
impurity Green function Eq. (\ref{eq:ImpurityGFApproximate}), the
effective polaron decay rate is actually given by $-\textrm{Im}\Sigma_{11}$
at $\omega=\mathcal{E}_{P}+\sqrt{\mathcal{Z}}\Omega/2$ (see, i.e.,
Eq. (\ref{eq:EnergyRabi})), which should increase with increasing
Rabi coupling. As a result of the additive effects of the momentum
average and Rabi coupling on enhancing the damping rate of Rabi oscillations,
we conclude that for attractive Fermi polarons, the damping rate of
Rabi oscillations can not be simply interpreted as the zero-momentum
quasiparticle decay rate $\Gamma$.

We finally note that the periodicity of Rabi oscillations is also
slightly affected by a finite Rabi coupling. Large Rabi coupling tends
to decrease and increase the periodicity for the repulsive polaron
and attractive polaron, respectively. It seems to have the same effect
as the momentum average, as shown in Fig. \ref{fig8_expt}. As a result,
for repulsive polarons, the combined additive effect of a finite Rabi
coupling and momentum average may lead to a smaller periodicity of
Rabi oscillations and hence a \emph{larger} effective Rabi coupling
$\Omega_{\textrm{eff}}$, compared with the expectation from a zero-momentum
Fermi polaron, i.e., $\sqrt{\mathcal{Z}}\Omega$. 

\section{Conclusions and outlooks}

In summary, based on the non-self-consistent many-body $T$-matrix
approximation, we have presented a general theoretical framework of
Fermi spin polarons for a spinor impurity immersed in a Fermi bath.
We have focused on the spin-1/2 case with a Rabi coupling $\Omega$
between the two spin states, and have addressed the dependence of
quasiparticle properties on the Rabi coupling strength. This turns
out to be crucial to understand the recent cold-atom experiments on
the dissipative Rabi dynamics of Fermi polarons \cite{Kohstall2012,Scazza2017,DarkwahOppong2019}.
In particular, we have confirmed that for the Rabi coupling less than
the Fermi energy of the Fermi bath, an approximate impurity Green
function provides a reasonable good description of Fermi spin polarons,
near the resonant detuning for the repulsive branch.

We have then developed an approximate theory for calculating the time-evolution
of the spin-down occupation, which is measured in the experiments
\cite{Kohstall2012,Scazza2017,DarkwahOppong2019}. This approximate
theory relies on a first order, mean-field type decoupling of a correlation
function that involves four field operators, which could be accurate
for the short-time evolution. We have compared our theoretical predictions
on Rabi oscillations with the experimental data \cite{Scazza2017}
and find a good agreement without any adjustable free parameters.
We have analyzed in detail the role played by the momentum average
on Rabi oscillations, due to the initial thermal distribution of the
impurity at finite temperature. We have also addressed the consequence
of a finite Rabi coupling at the order of the Fermi energy ($\Omega\sim\varepsilon_{F}$),
which could be significant in real experiments. The effects of both
factors (i.e., the thermal momentum average and the finite Rabi coupling)
are less considered in previous analyses of the dissipative Rabi dynamics
\cite{DarkwahOppong2019,Adlong2020}. 

We have found that, for repulsive polarons, the momentum average and
the finite Rabi coupling have opposite effects on the damping of Rabi
oscillations. As a result, to a good approximation, we may directly
extract the decay rate $\Gamma$ of a zero-momentum repulsive Fermi
polaron at zero Rabi coupling from the damping of Rabi oscillations,
a procedure that has already been experimentally adopted \cite{Scazza2017,DarkwahOppong2019}.
For the periodicity of Rabi oscillations, however, the two factors
have the same effects: both of them tend to decrease the periodicity
and hence lead to a slightly larger effective Rabi coupling strength
than the naive theoretical expectation of $\sqrt{\mathcal{Z}}\Omega$.

For attractive polarons, on the other hand, the situation turns out
to be more complicated. We have emphasized that at low temperature,
the zero-momentum decay rate of attractive Fermi polarons is not related
to the damping of Rabi oscillations. Both the thermal momentum average
and the finite Rabi coupling should be carefully taken into account
in analyzing the dissipative Rabi dynamics of attractive Fermi polarons.

We finally comment on the theoretical calculation of Rabi oscillations.
To go beyond the approximation of the mean-field decoupling, in Eq.
(\ref{eq:nt4f}) we may consider inserting the unity identify between
the field operators $d_{\mathbf{p}\downarrow}^{\dagger}(t)$ and $d_{\mathbf{p}\downarrow}(t)$,
\begin{equation}
1=\left|\textrm{FS}\left\rangle \right\langle \textrm{FS}\right|+\sum_{\mathbf{p},\mathbf{h}}c_{\mathbf{p}}^{\dagger}c_{\mathbf{h}}\left|\textrm{FS}\left\rangle \right\langle \textrm{FS}\right|c_{\mathbf{h}}^{\dagger}c_{\mathbf{p}}+\cdots,
\end{equation}
where the second term stands for the many-body state of the Fermi
bath with one-particle-hole excitations, and the terms in the ``$\cdots$''
denote the many-body states with multiple particle-hole excitations.
It is easy to see that the first term in the unity identity $\left|\textrm{FS}\left\rangle \right\langle \textrm{FS}\right|$
gives rise to the mean-field decoupling. The second term generates
the contributions that involve a correlation function,
\begin{equation}
\left\langle \textrm{FS}\left|d_{\mathbf{k}\downarrow}d_{\mathbf{p}\downarrow}^{\dagger}\left(t\right)c_{\mathbf{q}}^{\dagger}c_{\mathbf{p}+\mathbf{q}-\mathbf{k}}\right|\textrm{FS}\right\rangle .
\end{equation}
We will consider the calculation of this correlation function in future
works, with which we may recover the variational results presented
in Ref. \cite{Adlong2020}.
\begin{acknowledgments}
This research was supported by the Australian Research Council's (ARC)
Discovery Program, Grants Nos. DP240101590 (H.H.) and DP240100248
(X.-J.L.). Xia-Ji Liu was also supported in part by the National Science
Foundation under Grant No. PHY-1748958, during her participation in
the KITP program ``Living Near Unitarity'' and the KITP conference
``Opportunities and Challenges in Few-Body Physics: Unitarity and
Beyond''.

\textit{Note added}. --- After the submission of this manuscript,
we were informed by Tomasz Wasak their interesting related theoretical
work \cite{Wasak2022}, in which the decoherence and momentum relaxation
in Fermi-polaron Rabi dynamics have been analyzed by a kinetic equation
approach. An excellent agreement between their theoretical predictions
and the LENS experimental data is also demonstrated, without any free
parameters. The connection and comparison between our work and their
theoretical analysis will be addressed in future studies. We note
also that, most recently a strongly driven Fermi polaron has been
experimentally realized \cite{Vivannco2023}, which motivates us to
develop a more accurate theory of Fermi spin polarons beyond the many-body
$T$-matrix approximation and a better description of Rabi dynamics
than the current mean-field decoupling approach.
\end{acknowledgments}

\appendix

\begin{figure*}
\begin{centering}
\includegraphics[width=0.8\textwidth]{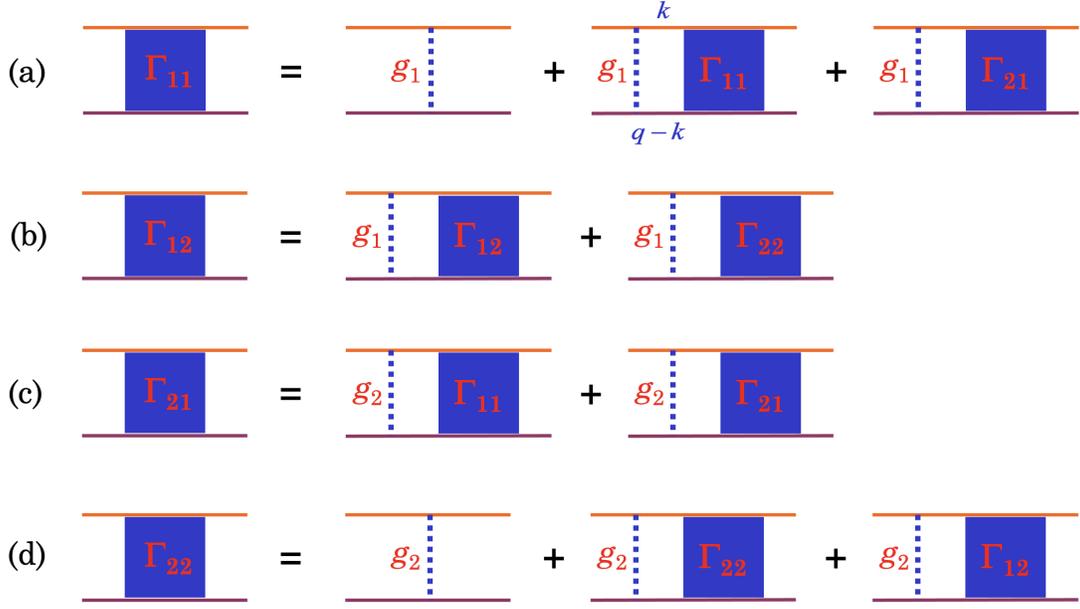}
\par\end{centering}
\caption{\label{figS1_diagram} Diagrammatic representation of the various
two-particle vertex functions $\Gamma_{ij}(\mathcal{Q})$. Here, the
upper orange line is the Green function of fermionic atoms in the
bath, and the bottom purple line is the impurity Green function. The
dotted line represents the contact interactions with strengths $g_{i}$
($i=1,2)$. We do not label explicitly the two hyperfine states of
the impurity.}
\end{figure*}

\section{The two-particle vertex function within ladder approximation}

The diagrammatic representation of the two-particle vertex functions
is shown in Fig. \ref{figS1_diagram}. For $\Gamma_{11}(\mathcal{Q})$
in Fig. \ref{figS1_diagram}(a) and $\Gamma_{21}(\mathcal{Q})$ in
Fig. \ref{figS1_diagram}(c), we can write down,
\begin{eqnarray}
\Gamma_{11}\left(\mathcal{Q}\right) & = & g_{1}-g_{1}\tilde{\chi}_{11}\left(Q\right)\Gamma_{11}-g_{1}\tilde{\chi}_{12}\left(Q\right)\Gamma_{21},\\
\Gamma_{21}\left(\mathcal{Q}\right) & = & -g_{2}\tilde{\chi}_{21}\left(Q\right)\Gamma_{11}-g_{2}\tilde{\chi}_{22}\left(Q\right)\Gamma_{21},
\end{eqnarray}
where the pair propagators $\tilde{\chi}_{ij}$ ($i,j=1,2$) are defined
in Eq. (\ref{eq:kappa2p}). By solving these two equations, we find
that,
\begin{eqnarray}
\Gamma_{11}\left(\mathcal{Q}\right) & = & \frac{\left(1/g_{2}+\tilde{\chi}_{22}\right)}{\left(1/g_{1}+\tilde{\chi}_{11}\right)\left(1/g_{2}+\tilde{\chi}_{22}\right)-\tilde{\chi}_{12}\tilde{\chi}_{21}},\\
\Gamma_{21}\left(\mathcal{Q}\right) & = & \frac{-\tilde{\chi}_{21}}{\left(1/g_{1}+\tilde{\chi}_{11}\right)\left(1/g_{2}+\tilde{\chi}_{22}\right)-\tilde{\chi}_{12}\tilde{\chi}_{21}}.
\end{eqnarray}
We similarly solve $\Gamma_{12}(\mathcal{Q})$ and $\Gamma_{22}(\mathcal{Q})$
by using the diagrams in Fig. \ref{figS1_diagram}(b) and Fig. \ref{figS1_diagram}(d).
It is readily seen that the final expressions for the various two-particle
vertex functions can be written in a compact form, as given by Eq.
(\ref{eq:vertexfunction}).

\end{document}